\documentclass[10pt]{article}

\usepackage{amsmath,amsbsy,amssymb}
\usepackage{upgreek,color}
\usepackage{graphicx}
\usepackage{float}
\usepackage{subfig}
\usepackage{caption}
\usepackage[]{geometry}
\usepackage{authblk}
\usepackage{natbib}

\usepackage{pstricks}

\usepackage[hidelinks,
            colorlinks=True,
            allcolors=black,
			citecolor=black]{hyperref}

\graphicspath{{figures/}{figures/bulle/}{figures/bulle_stat/}{images/}}

\newcommand \LADYF{{\normalsize L\kern -.35em\lower -.5ex\hbox{\scriptsize A} \normalsize\kern -.4em\lower .2ex
	\hbox{\footnotesize D}\kern -.25em\lower .ex\hbox{Y}\kern -.1em 
	\hbox{\footnotesize F}}}

\newcommand\kp{\ensuremath{\kappa^t_{t_0}}}
\newcommand\ko{\ensuremath{\kappa_0}}
\newcommand\kpm{\ensuremath{\overline{\kappa}^t_{t_0}}}
\newcommand\kpmT{\ensuremath{\overline{\kappa}^{t_0+T}_{t_0}}}
\newcommand\kpp{\ensuremath{\dot{\kappa}_{t_0}}}
\newcommand\fm{{\mathbf{F}}^t_{t_0}}
\newcommand\cg{\mathbf{C}^t_{t_0}}
\newcommand\rot{\ensuremath{\mathbf{R}}}
\newcommand\xb{\ensuremath{\mathbf{x}}}

\newcommand\rb{\ensuremath{\mathbf{r}}}

\newcommand\Sb{\ensuremath{\mathbf{S}}}
\newcommand{\dif}{\mathrm{d}}
\newcommand\B{\ensuremath{\mathcal{B}}}
\newcommand\Bo{\ensuremath{\B(t_0)}}

\newcommand\BEo{\ensuremath{\B_E(t_0)}}

\newcommand\Rey{\mbox{\textit{Re}}}  

\newcommand\pa{(\textit{a})}
\newcommand\pb{(\textit{b})}
\newcommand\pc{(\textit{c})}
\newcommand\pd{(\textit{d})}

\newcommand\pab{(\textit{a,\,b})}
\newcommand\pbc{(\textit{b,\,c})}

\newcommand{\fig}[1]{Fig.~\ref{fig#1}}
\newcommand{\figs}[1]{Figs.~\ref{fig#1}}
\newcommand{\Fig}[1]{Figure~\ref{fig#1}}
\newcommand{\Figs}[1]{Figures~\ref{fig#1}}

\begin{document}

\title{Material spike formation in high Reynolds number flow separation}
\author[1]{Mattia Serra%
	\thanks{Email address for correspondence: serram@seas.harvard.edu}}
\author[2]{Se\'an Crouzat}
\author[2]{Ga\"el Simon}
\author[2]{J\'er\^ome V\'etel%
	\thanks{jerome.vetel@polymtl.ca}}
\author[3]{George Haller\thanks{georgehaller@ethz.ch}\vspace{20pt}}
\affil[1]{School of Engineering and Applied Sciences, Harvard University,
Cambridge, MA 02138, USA\vspace{5pt}}
\affil[2]{Department of Mechanical Engineering, \LADYF, Polytechnique Montr\'eal, Montr\'eal, Qu\'ebec H3C\,3A7, Canada\vspace{5pt}}
\affil[3]{Institute for Mechanical Systems, ETH Z\"urich, 8092 Z\"urich, Switzerland}

\maketitle

\begin{abstract}
We apply the recent frame-invariant theory of separation spike formation to complex unsteady flows including a turbulent separation bubble, an impinging jet, and flows around a freely moving cylinder and a freely rotating ellipse. We show how the theory captures the onset of material spike formation, without any assumption on the flow type (steady, periodic, unsteady) or separation type (on- or off-wall, fixed or moving boundaries). We uncover new phenomena, such as the transition from on-wall to off-wall separation, the merger of initially distinct spikes, and the presence of severe material spikes that remain hidden to previous approaches. These results unveil how an involved network of spikes arises, interacts, and merges dynamically, leading to the final ejection of particles from the wall in highly transient flow separation processes. 
\end{abstract}

\section{Introduction}
Flow separation is important in engineering problems involving internal flows, such as diffusers \citep{Azad1996}, turbomachines and gas turbines \citep[see the review of][]{Cloos2017}, compressors \citep{Gresh2018}, heat exchangers \citep{Shah2003} or combustors \citep{Tianyun2017}, and external flows, such as those around airfoils and buildings \citep{Cermak1976}. Boundary layer separation increases the drag of surfaces in contact with fluids and often triggers the transition to turbulence. It may also generate critical undesirable phenomena, such as aerodynamic stall or compressor surge \citep[see the recent review of][]{Corke2015}. The prediction of separation phenomena, therefore, is crucial for air, land and marine transport, and for energy production. In the hydropower sector, the unsteady separation of swirling boundary layers is responsible for the degradation of turbines' performance \citep{Duquesne2015}. In aerodynamics, the peak efficiency is often reached close to the onset of separation which in turns also limit its performance. As a consequence, separation control strategies are in high demand.

There has been abundant literature on flow separation since the pioneering work of \cite{Prandtl1904} on two-dimensional steady flows, objective here is not to cover the whole subject, as pertinent reviews already exist \citep[see, e.g.,][]{Cassel2014}, but, instead, to recall the main approaches currently used. The vast majority of studies on unsteady separation has focused on the detection of a singularity in the boundary layer equation \citep{Sears1975,VanDommelen1982}. As examples of both separation without such singularities and singularities without separation are known \citep{liu1985simple}, this view practically associates separation with ones inability to solve the boundary-layer equations accurately. While the triple-deck theory \citep{Ruban2011} has partially solved the above limitation, all boundary-layer-singularity techniques are valid in the limit of infinite Reynolds number, as opposed to finite Reynolds number flows arising in practice.

Using dynamical systems theory, \cite{shariff1991dynamical} and \cite{yuster1997invariant} proposed a rigorous criterion for the existence of a material-spike on a no-slip boundary in near-steady, time-periodic incompressible flows. They defined the separation profile in such flows as the unstable manifold of a non-hyperbolic fixed point on the wall. Extending this idea, \cite{haller2004exact} developed a general theory of separation for a broader class of unsteady flows, defining two types of separation: fixed and moving separation. Fixed separation occurs in flows with a well-defined asymptotic mean \citep{kilic2005unsteady}, such as periodic and quasiperiodic flows, as well as aperiodic flows with a mean component. In this case, the separation point on the boundary is fixed at a location where the backward-time average of the skin friction vanishes; the angle of separation is generally time dependent. These results have also been extended to three-dimensional flows \citep{surana2006exact,surana2008exact}.

While the above approaches are radically different in some aspects, they nevertheless share a common feature: they are all based on asymptotic methods, targeting a separation profile with which particles ejected from the wall align in the limit of infinite time. Indeed, studies based on the boundary layer equation use asymptotic expansions to detect a thickening of the boundary layer or a large normal velocity component, whereas the dynamical systems theories are based on the backward-time asymptotic alignment of material lines with the wall. In flow control, however, the objective is to suppress flow separation at its onset rather than controlling it asymptotically. This requires a different approach that captures material spike formation over short time intervals, irrespective of the time scales and time dependence of the flow.



Separation is visualized in all experiments by observing the ejection of particles from a wall, such as Prandtl's aluminum foil in his original experiment. By a closer inspection of these experiments, one observes that the material ejection of particles from the wall is preceded by a sharp folding of the ejected fluid patch into a wall-transverse spike. Therefore, separation can be described as a material phenomenon where a layer of fluid undergoes a spike-shaped deformation before its ejection into regions far from a boundary. Based on this observation, \cite{Serra2018} studied the formation of such material spikes from the curvature evolution of material lines initially parallel to a no-slip boundary, and identified the spikes from the emergence of curvature maxima near the wall. In practice, the theory provides explicit frame-invariant formulae defining the material spike in two-dimensional, compressible or incompressible flows with arbitrary time dependence. It allows to detect the spike formation over very short-time intervals, or even instantaneously, contrary to previous criteria that seek to capture the long-term (asymptotic) behavior. As a consequence, the curvature based theory uncovers both on-wall and off-wall separation without a-priori assumptions on the flow.

In \cite{Serra2018}, however, only a few test cases were treated to illustrate the theory. Most examples invoked the analytic solution of the separated flow induced by the rotation and translation of a solid cylinder close to a wall at low Reynolds numbers \citep{Klono2001,Miron2015b}. A similar case, obtained experimentally, was also treated, and finally the numerical case of the flow over a circular cylinder was explored to validate the method for curved boundaries. Although the method successfully captured separation in all cases, these examples were limited to small Reynolds numbers. 

The objective of this study is to confirm the validity of the material spike formation theory to more challenging flows characterized by high Reynolds numbers. The paper is organized as follows. In \S\,\ref{th}, we summarize the relevant theoretical results from \cite{Serra2018} needed for our analysis. We then apply the spike formation theory to different test cases, including a separation bubble (\S\,\ref{bub}), an impinging jet (\S\,\ref{jet}), and flows around a freely translating cylinder and a freely rotating ellipse (\S\,\ref{fac}).

\section{Main results from the theory}\label{th}

Denoting by $\rb: s \mapsto \rb (s), s \in [s_1,s_2] \subset \mathbb{R}$, the arclength parametrization\footnote{The theory of \cite{Serra2018} is valid for arbitrary parametrization, but here we adopt arclength parametrization for simplicity.} of a material curve $\gamma \subset D\subset \mathbb{R}^2$, and by $(\cdot)'$ differentiation with respect to the curvilinear coordinate $s$, the theory described in \cite{Serra2018} provides the exact Lagrangian curvature evolution of $\gamma$ from its initial curvature $\kappa_0 $ to $\kappa^t_{t_0}$ between the initial time $t_0$ to the current time $t$ for any time-dependent flow. For incompressible flows, which will be of interest in this article, one obtains
\begin{align}
\kp = \frac{\left< (\nabla^2 \fm(\rb)\rb')\rb', \rot\nabla\fm(\rb)\rb'\right>}{\left<\rb',\cg(\rb)\rb'\right>^{3/2}}
+ \frac{\ko}{\left<\rb',\cg(\rb)\rb'\right>^{3/2}},
\label{k1}
\end{align}
where $\langle \cdot , \cdot \rangle$ denotes the inner product; $(\nabla \fm(\rb)\rb^\prime)_{ij}=\sum\limits_{k=1}^{2}{\partial_ {jk}F_{t_{0}}^t}_{i}(\rb)r_k^\prime,\ i,j\in\{1,2\}$, where $\fm$ is the flow map defined by the fluid trajectories:
\begin{align}
	\fm(\xb_0) = \xb_0 + \int_{t_{0}}^{t}\mathbf{v}( \mathbf{F}_{t_{0}}^\tau(\xb_0),\tau)\ d\tau;
\end{align}
$\cg =[\nabla \fm]^\top \nabla \fm$ is the right Cauchy--Green strain tensor, and \rot\ is the rotation matrix defined as
\begin{align}
	\rot := \begin{bmatrix}
0 & 1 \\
-1 & 0
\end{bmatrix}.
\end{align}\label{kt}

 \begin{figure}[h!]
	\centering
	\includegraphics[height=.13\columnwidth]{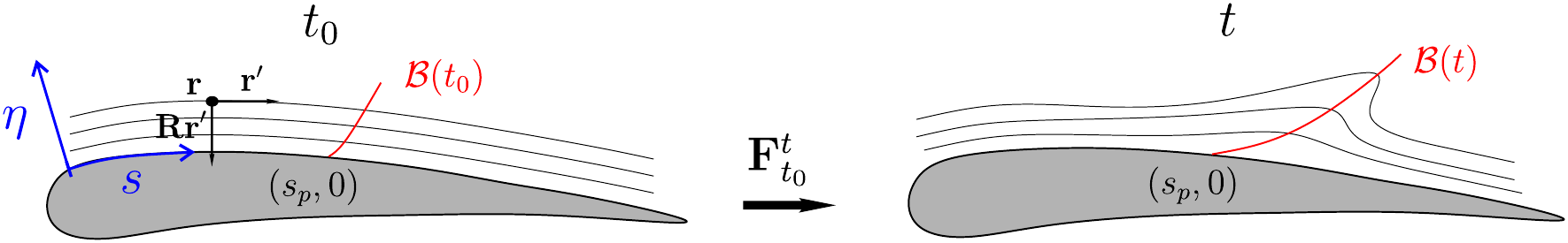}\label{fig:DeathByRescue}
	\caption{(left) Near-wall region foliated by material lines (black curves) initially parallel to the wall, and parametrized using the $[s,\eta]$ coordinates. The red curve shows the initial position $\B(0)$ of the Lagrangian backbone of separation whose intersection with the wall defines the Lagrangian spiking point $s_p$. (right) Advected material lines (black) along with the current position $\B(t)$ of backbone of separation acting as the theoretical centerpiece of the forming spike.}	
	\label{fig:illustr}
\end{figure}

Using Eq. \eqref{k1}, we compute the curvature change relative to the initial curvature $\kpm := \kappa^t_{t_0} - \kappa_0$ in a neighborhood of the no-slip boundary foliated by a set of material lines initially parallel to the wall, parametrized using the stream-wise $s$ and wall-normal $\eta$ coordinates (Fig. \ref{fig:illustr}). Such a foliation enslaves the initial local tangent $\rb'$ and curvature $\kappa_{0}$ to the position $\rb$, making therefore $\kpm$ a function of $t_{0},t$ and of the initial configuration $\rb$ only. The initial position $\mathcal{B}(t_0)$ of the \textit{Lagrangian backbone of separation} -- i.e., the theoretical centerpiece of the material spike over $[t_0,t]$ -- is then defined as a positive-valued wall-transverse ridge of the $\kpm$ field.  In other words, $\mathcal{B}(t_0)$ is the set of points where the curvature change attains a local maximum with respect to wall-parallel direction.
Later position of the backbone $\mathcal{B}(t)$ can be computed by materially advecting $\mathcal{B}(t_0)$, i.e., letting $\mathcal{B}(t) := \fm(\mathcal{B}(t_0))$. If $\mathcal{B}(t_0)$ connects to the wall, the separation is on-wall, otherwise off-wall. In the case of on-wall separation, the intersection of $\mathcal{B}(t_0)$ and the wall defines the \textit{Lagrangian spiking point} $s_p$, which captures the on-wall signature of the spike formation (Fig. \ref{fig:illustr}). Analytic expression for $s_p$ are available in \cite{Serra2018}.


In the limit of $t\rightarrow t_0$, the instantaneous limits of the Lagrangian backbone of separation and spiking point are the \textit{Eulerian backbone of separation}  $\mathcal{B}_E(t)$ and \textit{Eulerian spiking point} $s_{pE}$. These can be computed by noting that
\begin{align}\label{k3}
\kpp = \left. \frac{\dif \kpm }{\dif t} \right|_{t=t_0},
\end{align}
and following the procedure above, substituting $\kpm$ with $\kpp$.
\cite{Serra2018} provides an explicit formula for the curvature rate of a material line, which in the case of incompressible flows and arclength parametrization reads

\begin{equation}
\label{eq:kdot}
\dot{\kappa}_t = \left<\rot\rb', (\nabla \Sb(\rb,t)\rb')\rb' \right> - \frac{1}{2} \left< \nabla \omega (\rb,t),\rb' \right> -3 \kappa \left< \rb', \Sb(\rb,t)\rb'\right>,
\end{equation}
where $\Sb$ denotes the rate-of-strain tensor and $\omega$ the scalar vorticity of the underlying velocity field $\mathbf{v}(\mathbf{x},t)$. Equation \eqref{eq:kdot} indicates that the spike formation arises from an interplay of stretching and rotation, combined in the material curvature field in a frame-invariant fashion. Even though the vorticity is a frame-dependent quantity, its spatial gradient, which naturally enters in Eq. \eqref{eq:kdot}, is objective \citep{Serra2018}. 
\begin{figure}[H]
    \centerline{\includegraphics[scale=0.4]{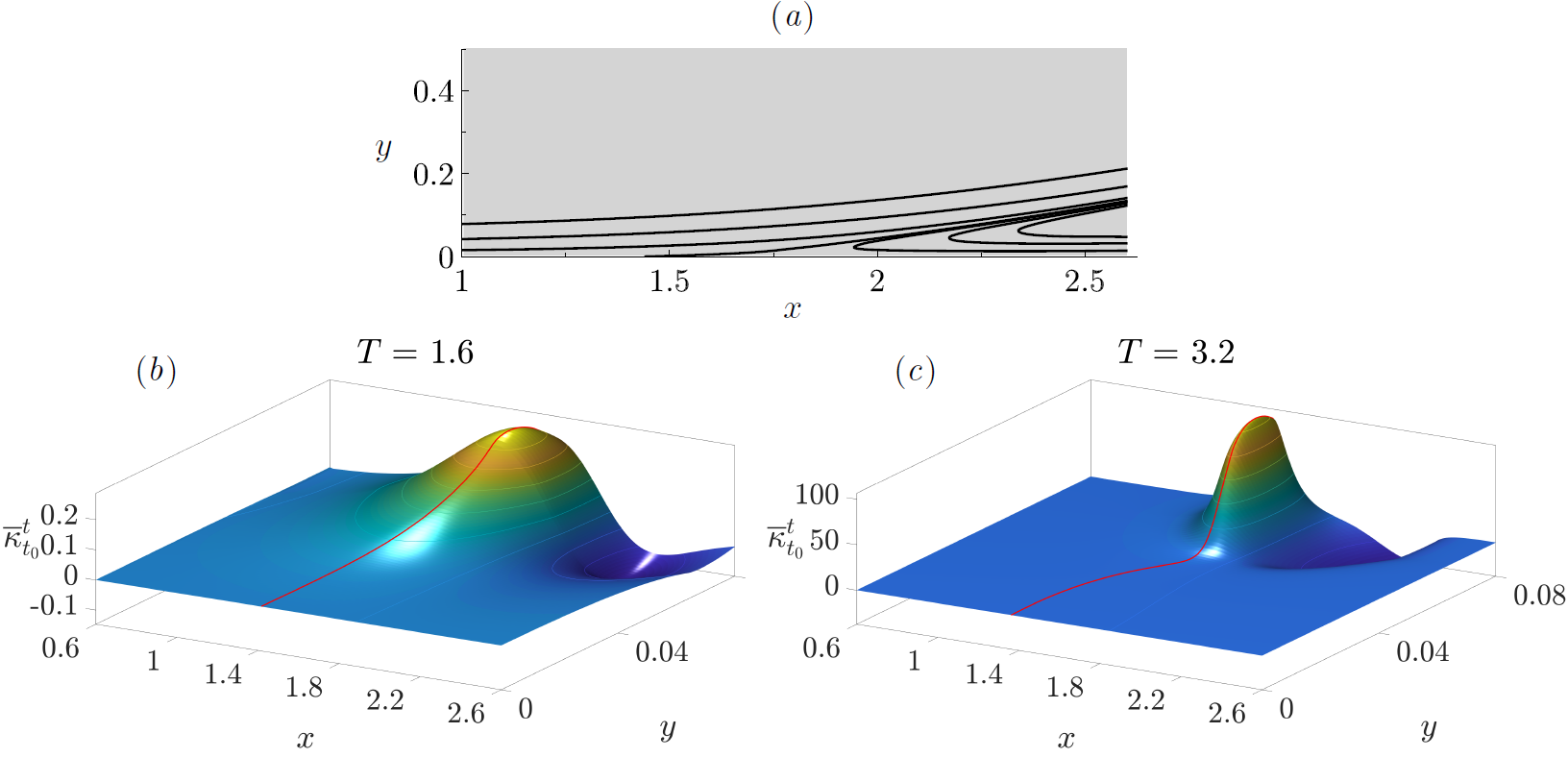}}
    \caption{\pa\ Example of streamlines in a laminar separation bubble flow (see \S\,\ref{bub} for details) and Lagrangian curvature change field $\kpmT$ for $T=1.6$ \pb\ and 3.2 \pc. The red curve shows the initial position of the Lagrangian backbone of separation $\B(0)$ defined as a wall-transverse ridge of $\kpmT$.}
    \label{fig1}
\end{figure}

As an illustrative example, \fig{1}\pa\ shows streamlines of a separated flow in the vicinity of the wall located at $y=0$ (see the laminar separation bubble described in \S\,\ref{bub} for details). \Fig{1}\pb\ shows the Lagrangian curvature change field \kpmT\ for the integration time $T=1.6$ along with its wall-transverse ridge $\mathcal{B}(t_0)$ in red. \Fig{1}\pc, shows the same as \Fig{1}\pb\ for a longer integration with $T = 3.2$. For longer integration time, the material spike is expected to be sharper, which is confirmed by the shape and magnitude of the \kpm\ field. \textcolor{black}{The flow in \fig{1} is slightly unsteady and hence Prandtl's criterion is technically inapplicable, Fig. \ref{fig1} shows that the Lagrangian spiking point is at $x\approx1.4$ while the zero skin friction point at $x\approx1.8$, highlighting that the Lagrangian backbone of separation reveals information that remain hidden to instantaneous streamlines.}
\begin{figure}[h]
    \centerline{\includegraphics[scale=.4]{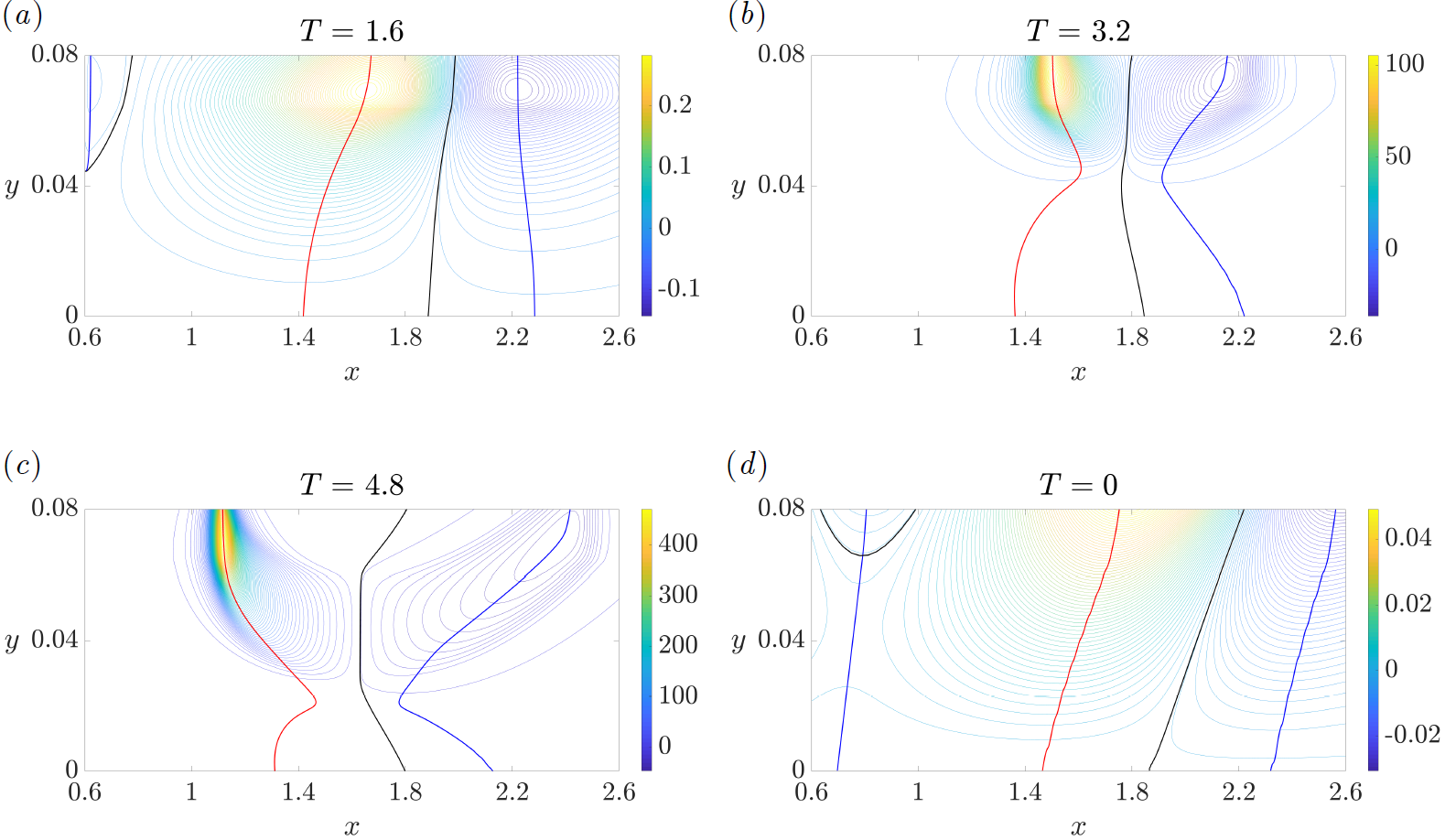}}
    \caption{\pab\ Contour plots of the scalar fields in \fig{1}\pbc, together with a longer time $T$ in \pc. \pd\ Contour plot of the curvature rate field $\kpp$ which corresponds to the time derivative of $\kpmT$ evaluated at $T=0$. The red curves show the initial position of the Lagrangian (resp. Eulerian) backbone of separation \Bo\ (resp. \BEo), the blue curves show the sets of minimal signed curvature (resp. curvature rate) in the vicinity of the backbone and the black curves represent the zero set of $\kpmT$ (resp. $\kpp$).}
    \label{fig2}
\end{figure}
To illustrate the shape of the initial positions of the Lagrangian backbones of separation \B(0) and its dependence on $T$, \figs{2}\pab\ show the contour plots of the \kpmT\ from \figs{1}\pbc. \Fig{2}\pc\ shows the same as the above panels for an even longer $T$, while \fig{2}\pd\ shows the curvature rate field \kpp\ that corresponds to the time derivative of \kpmT\ evaluated at $T=0$. Red lines show the Lagrangian and Eulerian backbones of separation. Blue lines show the loci of minimal signed curvature in their vicinity, while black lines show the zero level set of \kpmT. \Fig{2}\ shows that the shape of the backbone of separation strongly depends on $T$, while the spiking point remains almost at the same location. \cite{Serra2018} show that $s_p\equiv s_{pE}$, and is independent of $T$ when the flow is steady. Here, however, the flow is slightly unsteady, generating oscillations of $s_p$ as a function of $T$.

\begin{figure}[H]
    \centering
    \centerline{\includegraphics[scale=1.1]{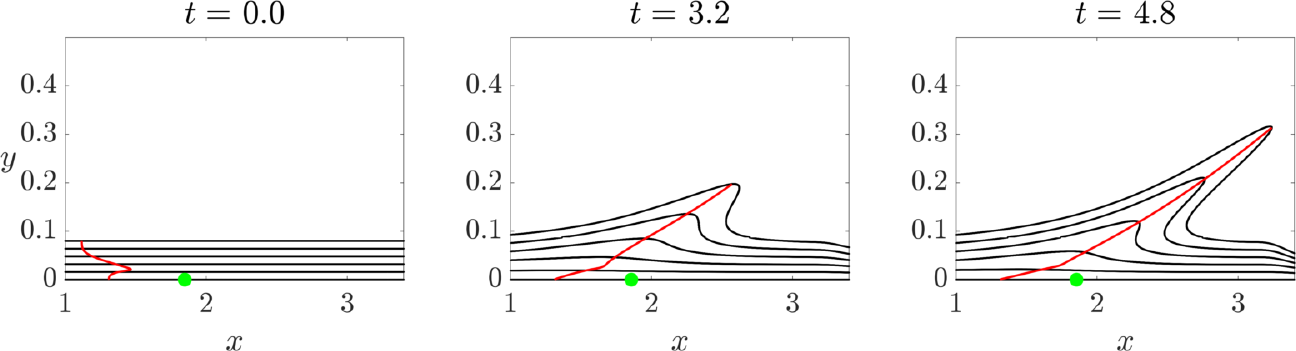}}
    \caption{Time evolution of the Lagrangian backbone $\B(t)$ of separation computed for $T = 4.8$, along with material lines initially parallel to the wall. The green dot represents the Prandtl separation point (zero wall shear point).} 
    \label{fig3}
\end{figure}


\Fig{3} shows the time evolution of the Lagrangian backbone of separation $\B(t)$ computed for $T = 4.8$, which acts as the theoretical centerpiece of the forming spike, along material lines initially parallel to the wall (black). The green dot represents the zero-skin-friction point, i.e. the Prandtl separation point. Note that the spiking point is at an upstream location compared to the Prandtl point even in steady flows \citep[see also][]{Serra2018}. 

\section{Separation bubble}\label{bub}
\subsection{Flow conditions}

Here we consider material spike formation in a two-dimensional (2D) separation bubble. The computational domain is $L_x  \times L_y = 8h\times 2h$ where $h$ is the channel half-height ($x$ and $y$ are the streamwise and normal coordinates non-dimensionalized by $h$). To generate a separated flow on the bottom no-slip wall ($y=0$), we prescribe a given velocity profile on the top wall ($y=2$) which induces an adverse pressure gradient. At the flow exit, we use a conventional convective outflow boundary condition. We use the 2D version of the finite difference code \texttt{Incompact3d} \citep{Laizet2009,Laizet2011} to solve the incompressible Navier--Stokes equations, and we show  in \fig{4}\pa\ an example of the vorticity field for this flow when a Blasius velocity profile is prescribed at the inlet. Two class of separation phenomena may occur. First, a quasi-steady separation appears at $x\sim 1.5$ due to the presence of an adverse pressure gradient, which is the case we showed as illustration in \S\,\ref{th}. Second, multiple vortex-induced separation phenomena can simultaneously occur in the wake of the instability triggered by the previous separation.

\begin{figure}[h!]
	\centerline{\includegraphics[scale=1]{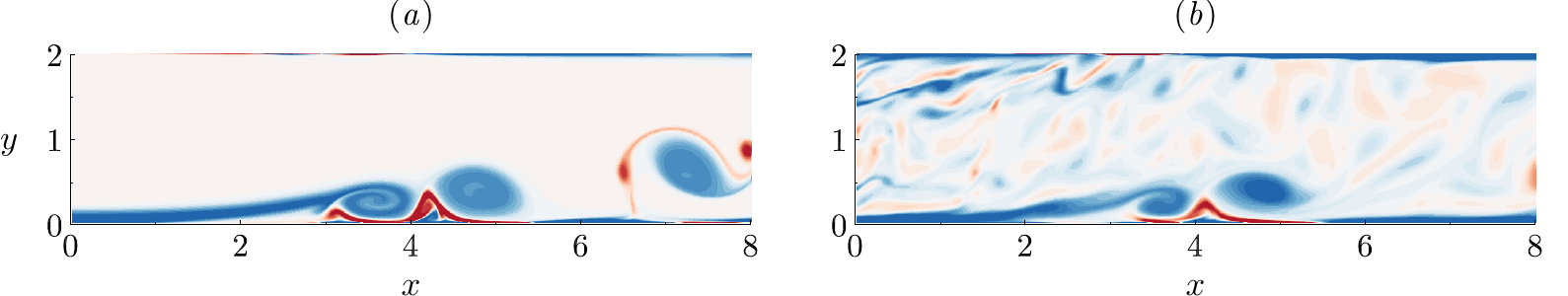}}
	\caption{Snapshot of the vorticity field for the laminar \pa\ and the turbulent \pb\ separation bubble.}
	\label{fig4}
\end{figure}

In order to study the spike formation in more complex flows, we impose a turbulent inlet flow condition. In this case, we solve the Navier--Stokes equations using the 3D version of the code \texttt{Incompact3d}, which provides a fully developed channel flow, periodic in the streamwise direction, at $\Rey=Uh/\nu = 5000$, where $U$ is the bulk velocity and $\nu$ the kinematic viscosity of the fluid. \textcolor{black}{After solving the 3D Navier-Stokes, we stored the resulting velocity field, and then used it as a time dependent inflow condition for the 2D computation of the separation bubble}. We call this flow a turbulent separation bubble, despite the flow being 2D, because of its chaotic behavior.

\subsection{Results}\label{sec:SBres}
\noindent With turbulent inflow conditions, the main separation described in \S\,\ref{th} is chaotic (see the vorticity field in \fig{4}\pb). \Fig{5}\pa\ shows the Lagrangian curvature field $\overline{\kappa}^6_0$ in the upstream part of the flow, while \fig{5}\pb\ shows the corresponding contour plot.
\begin{figure}[h!]
    \centerline{\includegraphics[scale=.4]{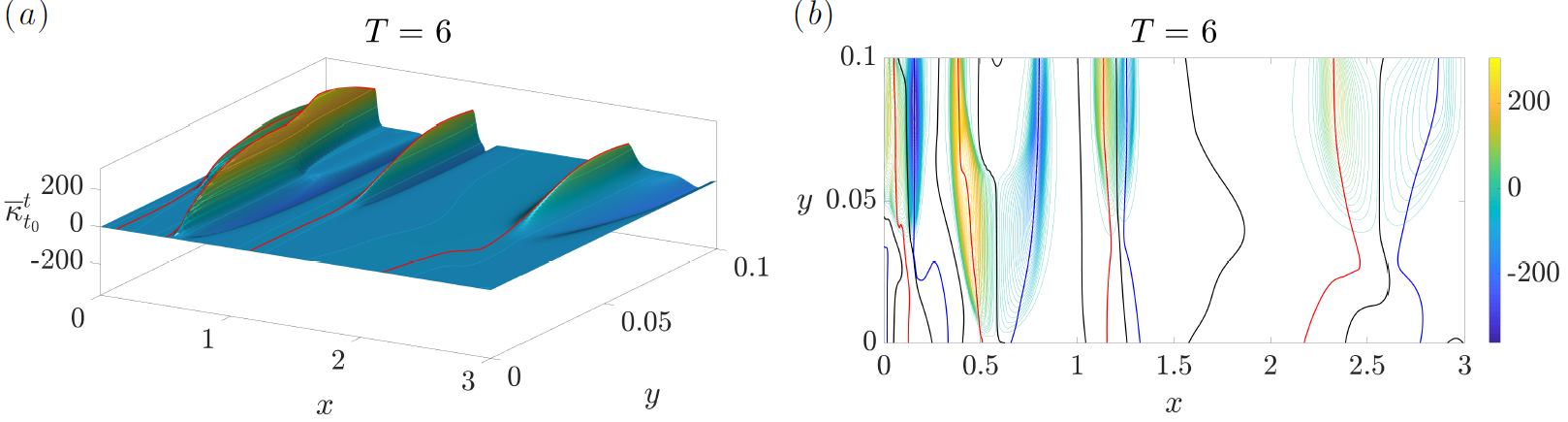}}
    \caption{\pa\ Lagrangian curvature change field $\overline{\kappa}^6_0$ in the vicinity of the first separation $(x < 3)$ in the turbulent separation bubble. \pb\ Contour plot of the scalar field in \pa.}
    \label{fig5}
\end{figure}
 The curvature topology is more complex than the one in the previous section, leading to four coexisting Lagrangian backbones of separation. Despite such complexity, each backbone is located between two zero set of $\kpm$ (black lines), and further by two minimal signed curvature (blue lines).
\begin{figure}
    \centerline {\includegraphics[scale=1]{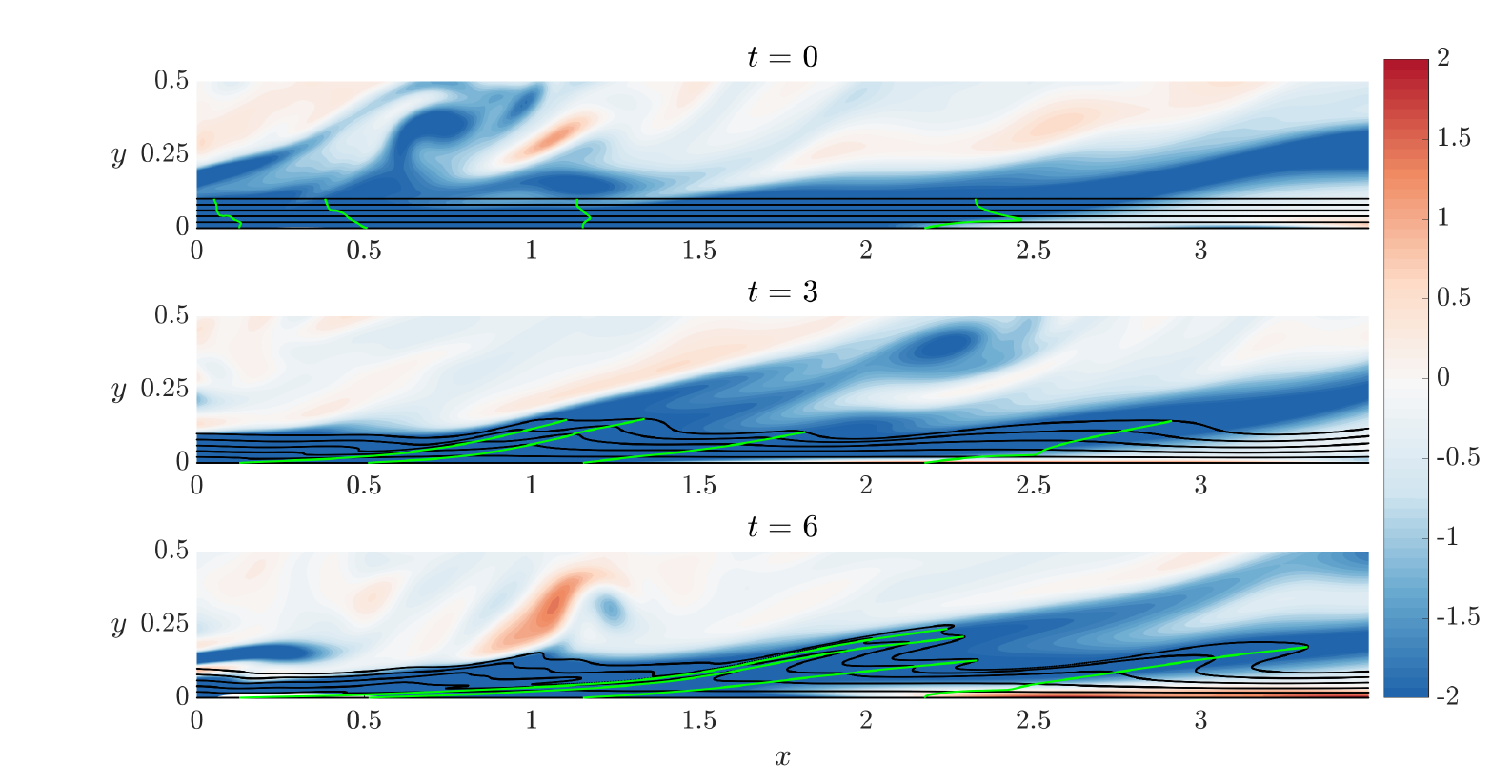}}
    \caption{Time evolution of the Lagrangian backbones of the separation $\B(t)$ (green) extracted from $\overline{\kappa}^6_0$, along with the vorticity field and material lines initially parallel to the wall (black) in the upstream region of the turbulent separation bubble.} 
    \label{fig6}
\end{figure}
\Fig{6} shows the time evolution of the Lagrangian backbones of separation \B($t$) (green) extracted from $\overline{\kappa}_0^6$ (cf. \fig{5}), along with vorticity field and material lines initially parallel to the wall (black). Material advection again shows that \B($t$) act as the backbones of forming spikes. Here, time $t$ is non-dimensionalized with $U$ and $h$. Despite the unsteadiness of the flow, all backbones have distinct on-wall footprints, thus revealing a complex network of separation process induced by small-scale vortices. 
\begin{figure}
    \centerline{\includegraphics[scale=.4]{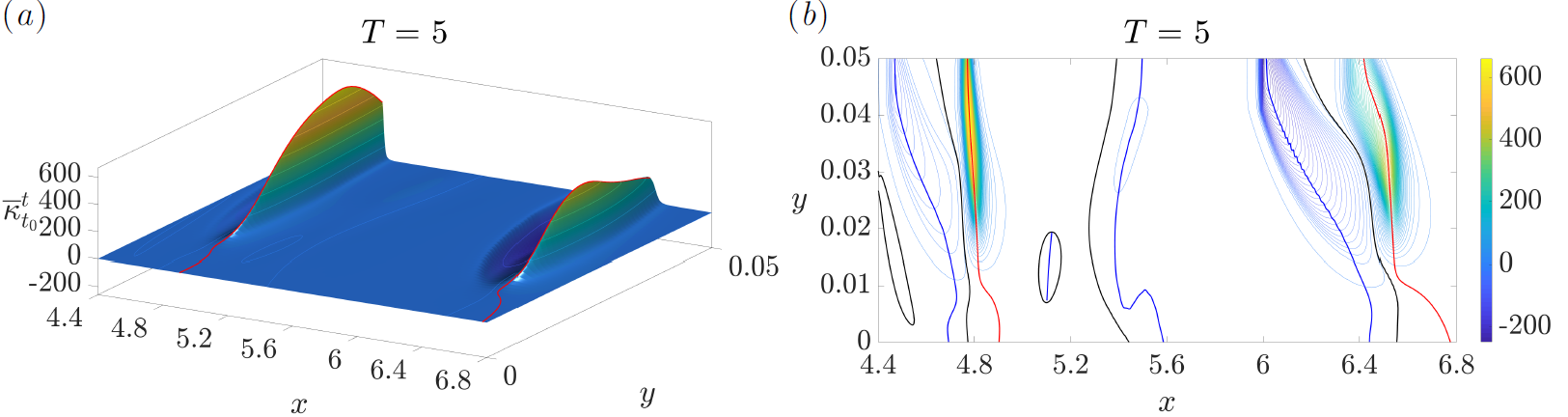}}
    \caption{\pa\ Lagrangian curvature change field $\overline{\kappa}^5_0$ for vortex-induced separations $(x \geq 3)$ in the turbulent separation bubble. \pb\ Contour plot of the scalar field in \pa.}
    \label{fig7}
\end{figure}

In the downstream part of the channel ($x \geq 3$), separation induced by large vortices dominates the flow dynamics (cf. \fig{4}\textit{b}). \Fig{7} shows the Lagrangian curvature field computed for $t_0 = 0$ and $T=5$, from which two backbones, connected to the wall, can be clearly identified. \Fig{8} shows these Lagrangian backbones extracted from $\overline{\kappa}_0^5$ at different times (green), along with the vorticity field and material lines initially parallel to the wall (black). It is clear from the figure that the entrainment induced by vortices $V_1$ and $V_2$, highlighted by negative vorticity contours (clockwise), for example at $t=1$, are responsible for the formation of the two material spikes identified in \fig{7}. 
\begin{figure}[H]
	\centerline{\includegraphics[scale=1]{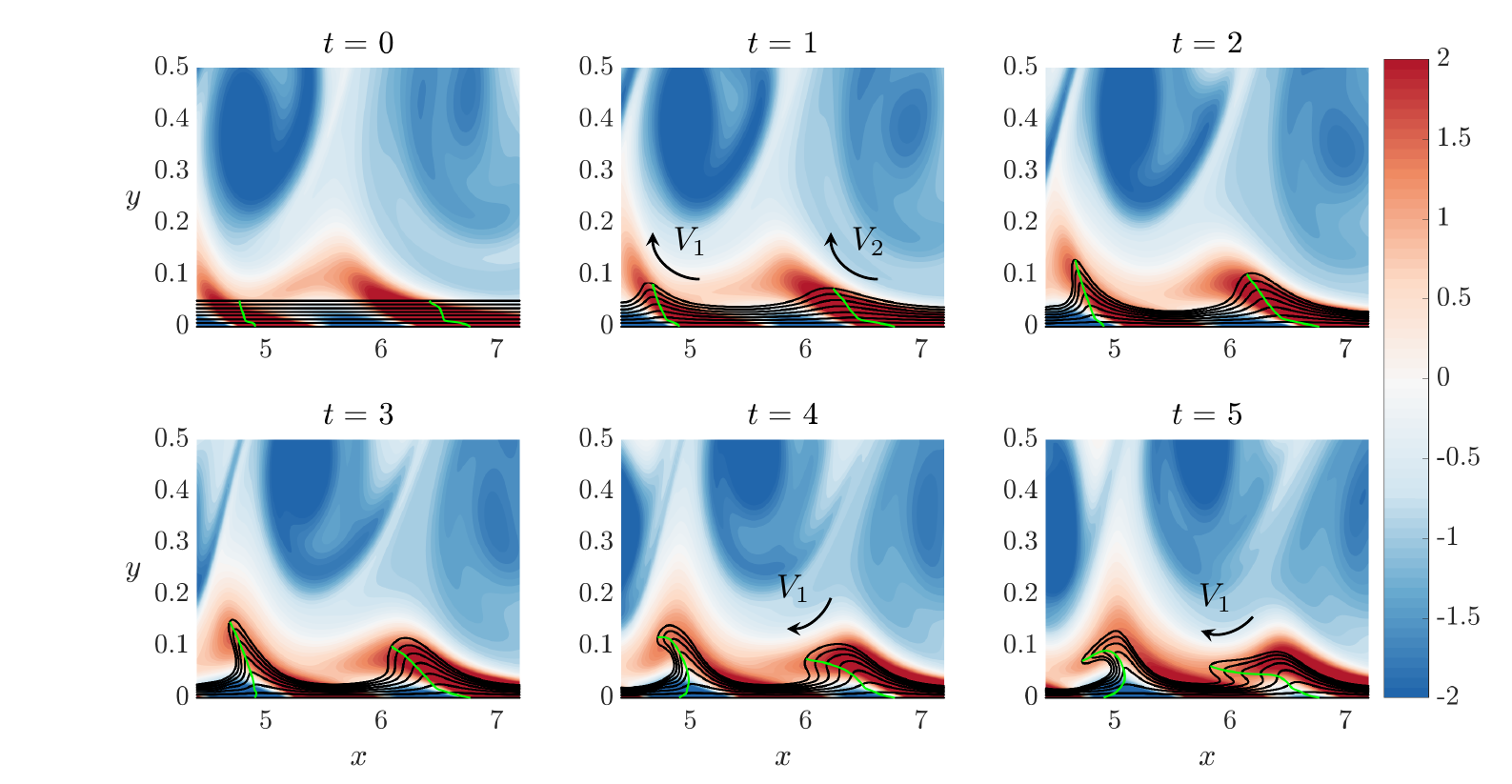}}
	\caption{Time evolution of the Lagrangian backbones $\B(t)$ of separation (green) computed from $\overline{\kappa}_0^5$, along with the vorticity field and material lines initially parallel to the wall (black) for vortex-induced separations in the turbulent separation bubble. Black arrows illustrate the fluid entrainment by vortices $V_1$ and $V_2$.} 
	\label{fig8}
\end{figure}
\noindent As $V_1$ convects to larger $x$, the induced velocity field on the downstream separation has progressively an opposite effect because the backbone is pushed back towards the wall ($t \geq 4$) due to the $V_1$ rotation direction, before being ejected again from the wall (not shown). By using the present theory, we can actually identify the presence of two separation backbones even in the instantaneous limit, i.e., as Eulerian backbones of separation, as shown in \fig{9}.

\begin{figure}[H]
    \centerline{\includegraphics[scale=.4]{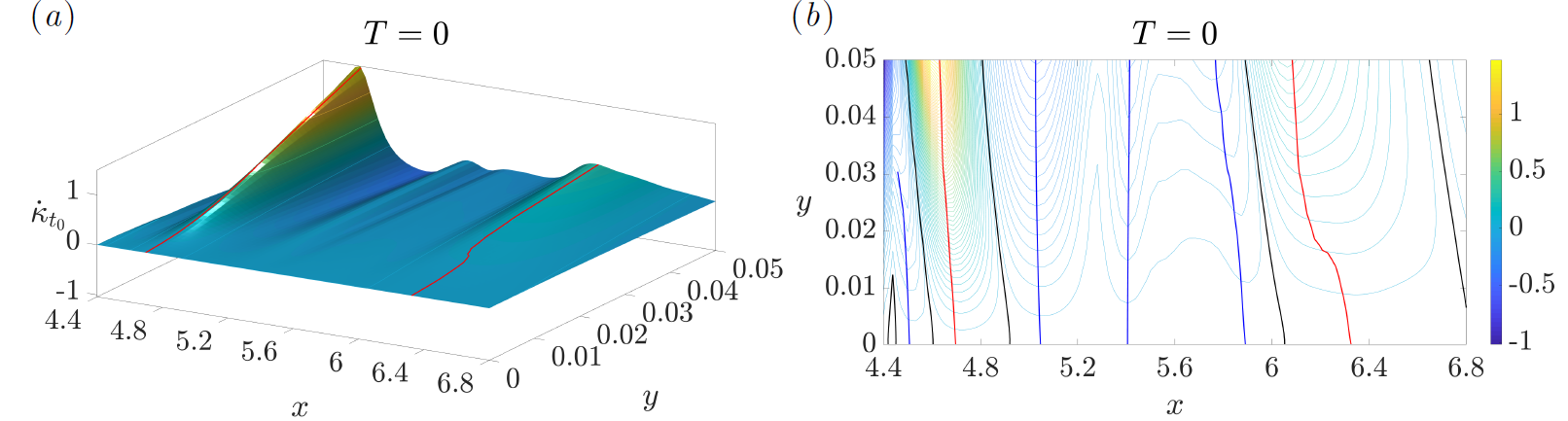}}
    \caption{\pa\ Eulerian curvature rate field $\dot{\kappa}_0$ for vortex-induced separations $(x \geq 3)$ in the turbulent separation bubble. \pb\ Contour plot of the scalar field in \pa.}
    \label{fig9}
\end{figure}

\section{Impinging jet}\label{jet}

\subsection{Flow conditions}
The impinging jet is a challenging test case for unsteady separation studies as various types of separation can occur, despite the simple geometry. We use the open source finite element code \texttt{FreeFem++} \citep{Freefem} to solve the incompressible Navier--Stokes equations for an axisymmetric impinging jet. The Reynolds number, based on the jet diameter $D$ and the jet velocity $U$, is $\Rey = 1000$. \Fig{10} shows an example of vorticity field. The jet is oriented from the top to bottom, with its symmetry axis located at $x=0$. The nozzle-to-plate distance is $4D$ ($x$ and $y$ are the radial and normal coordinates non-dimensionalized by $D$). We impose a hyperbolic tangent velocity profile at the jet exit velocity (grey velocity vectors in \fig{10}), and no-slip conditions at the upper and lower walls.

\begin{figure}[H]
	\centerline{\includegraphics[scale=0.8]{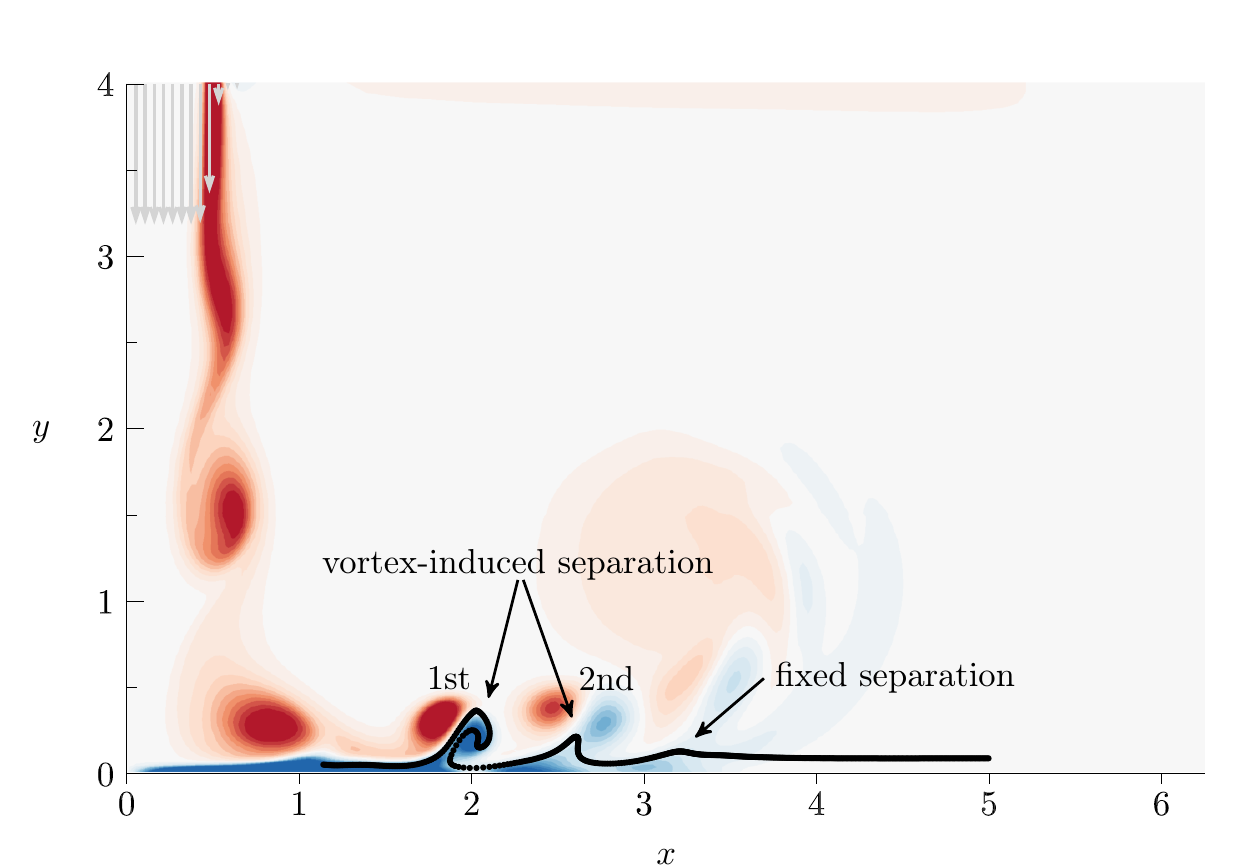}}
	\caption{Vorticity plot in the axisymmetric impinging jet flow (the jet axis is located at $x=0$). The jet is oriented from top to bottom, as indicated by velocity vectors (in grey) at the nozzle exit. Black points show positions of particles initially aligned with the wall and advected at an arbitrary time to illustrate different separation types.} 
	\label{fig10}
\end{figure}

When primary vortices (red vorticity in \fig{10}) forming in the jet mixing layer impact on the wall, they generate secondary counter-rotating vortices (blue) moving parallel to the bottom wall. This is accompanied by an unsteady separation that moves with the primary vortices \citep[see, e.g.,][]{Didden1985}. It is generally accepted that this separation process, that is \textit{moving}, is not connected to the wall. In \cite{Miron2015b} for example, the separation point attached to each vortex was defined \textit{a priori} as a saddle point \textit{off} the wall. Such a point can be detected by analyzing an exponent that cumulates the history of the strain rate along a repelling Lagrangian coherent structure (LCS), extracted, for example, as in \cite{Farazmand2012}, as a material line with the highest normal repulsion rate \citep{Haller2011}. The separation point can then be subsequently followed in time by material advection. This discussion is relevant for the first and second vortex-induced separation processes indicated in \fig{10}.

As suggested in \fig{10}, the first two separations are usually considered as moving, without any connection to the impinged wall. \cite{Lamarche2018} show that for a higher radial position, there is also \textit{fixed} separation (see \fig{10}). They find that such a fixed separation is observed where the one-period-averaged wall shear-stress vanishes, confirming the asymptotic theoretical results of \cite{haller2004exact} for periodic flows. \cite{Lamarche2018} also observed that that a material spike forms upstream of the fixed separation point, as already observed in \S\,\ref{th}. \textcolor{black}{
In summary, separation in the impinging jet flow has been studied using asymptotic methods \citep{haller2004exact, Miron2015b}, and assuming a priori whether separation should be on- or off- wall. Here, by contrast, we use the theory of \cite{Serra2018}, which is assumption-free and targets spike formation rather than asymptotic separation profiles.}

\subsection{Results}\label{sec:IJRes}
\Figs{11} and \ref{fig12} are related to the first vortex-induced separation in \fig{10}.
Specifically, \fig{11} shows the Lagrangian curvature change field obtained for $T=1$, where time is non-dimensionalized with $U$ and $D$, signaling two backbones for this separation. \Fig{12} shows the later position of $\B(t_0)$ from \figs{11}, along with material lines and the vorticity field, confirming that the first separation is indeed characterized by two different spikes that evolve closer over time. These results not only reveal that multiple spikes can contribute to the same long-time separation phenomenon, but most importantly, show that both backbones have a footprint on the wall, in disagreement with previous studies in which this kind of separation was treated a-priori as off-wall. We emphasize that not being able to find an on-wall signature of a separation phenomenon from a given approach does not imply that it indeed has none. As an advantage, the theory of \cite{Serra2018} is free from a priori assumptions on the separation type: whether a separation is on- or off-wall is a result, rather than an input, of the method.

\begin{figure}[H]
    \centerline{\includegraphics[scale=.35]{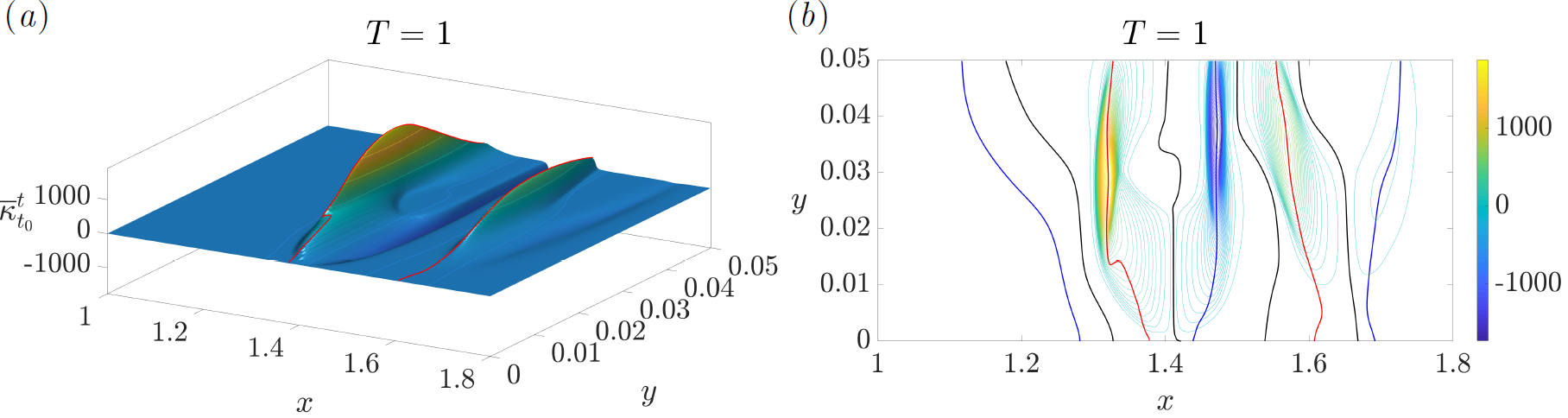}}
    \caption{\pa\ Lagrangian curvature change field $\kpmT$ for $T = 1$ for the first vortex-induced separation in the impinging jet (see \fig{10}). \pb\ Contour plot of the scalar field in \pa.}
    \label{fig11}
\end{figure}

\begin{figure}[H]
    \centerline{\includegraphics[scale=1]{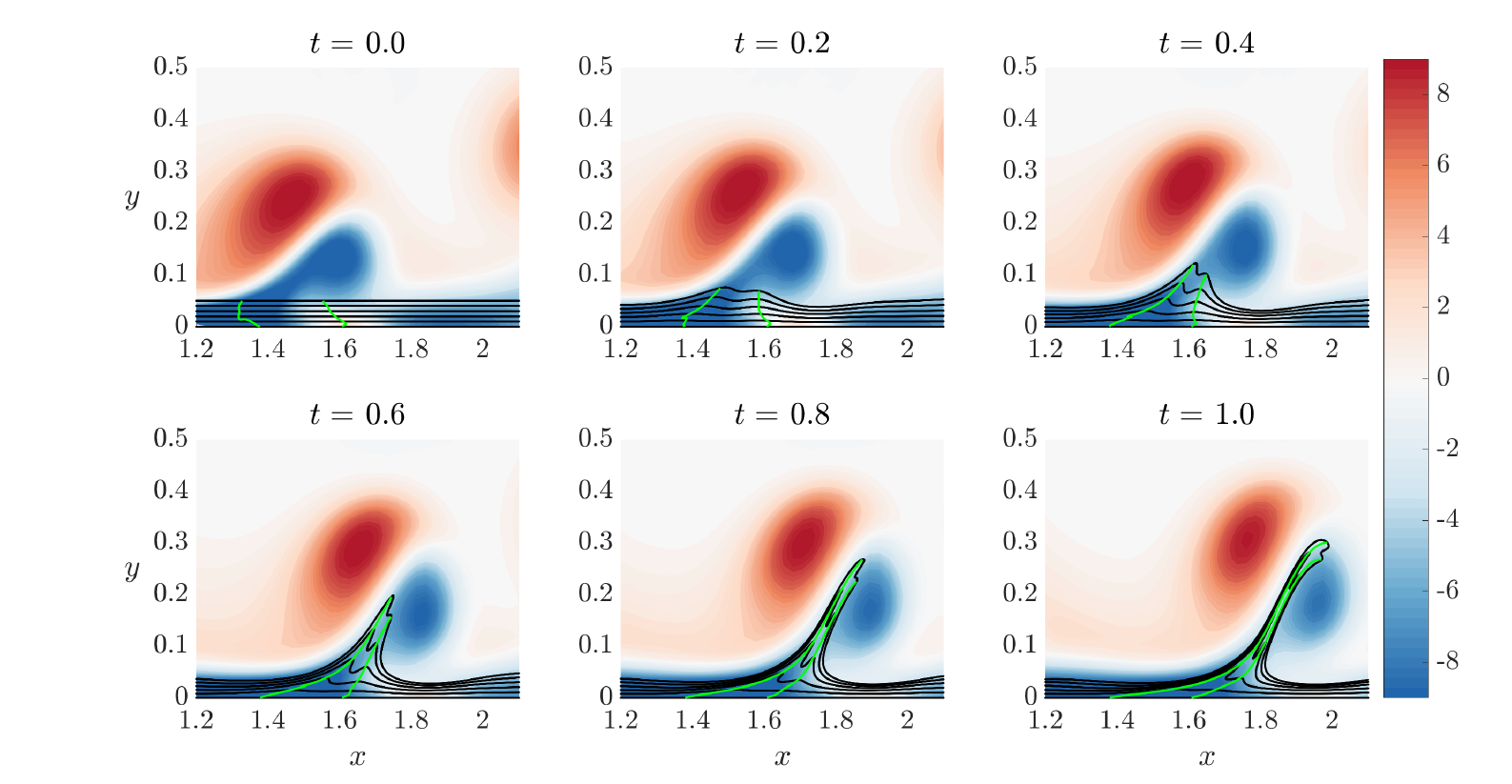}}
    \caption{Time evolution of the Lagrangian backbones of separation $\B(t)$ (green) computed from $\overline{\kappa}_0^1$, along with the vorticity field and material lines initially parallel to the wall (black) for the first vortex-induced separation in the impinging jet (see \fig{10}).} 
    \label{fig12}
\end{figure}

\begin{figure}[H]
    \centerline{\includegraphics[scale=.4]{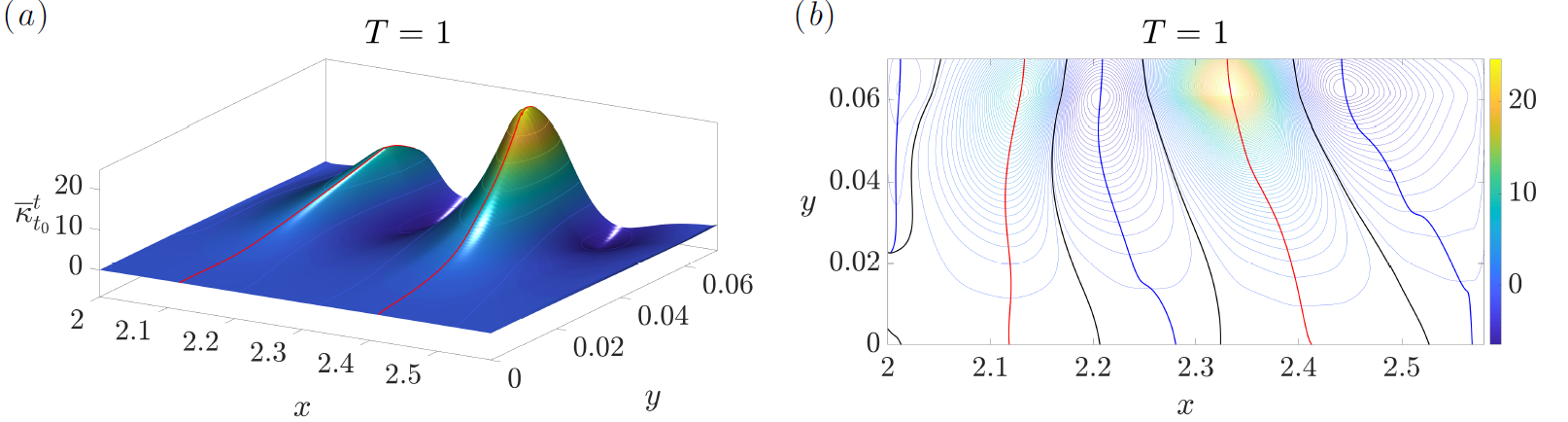}}
    \caption{\pa\ Lagrangian curvature change field $\overline{\kappa}^1_0$, in correspondence of the second vortex-induced separation in the impinging jet (see \fig{10}). \pb\ Contour plot of the scalar field in \pa.}
    \label{fig13}
\end{figure}

\begin{figure}[H]
    \centerline{\includegraphics[scale=1]{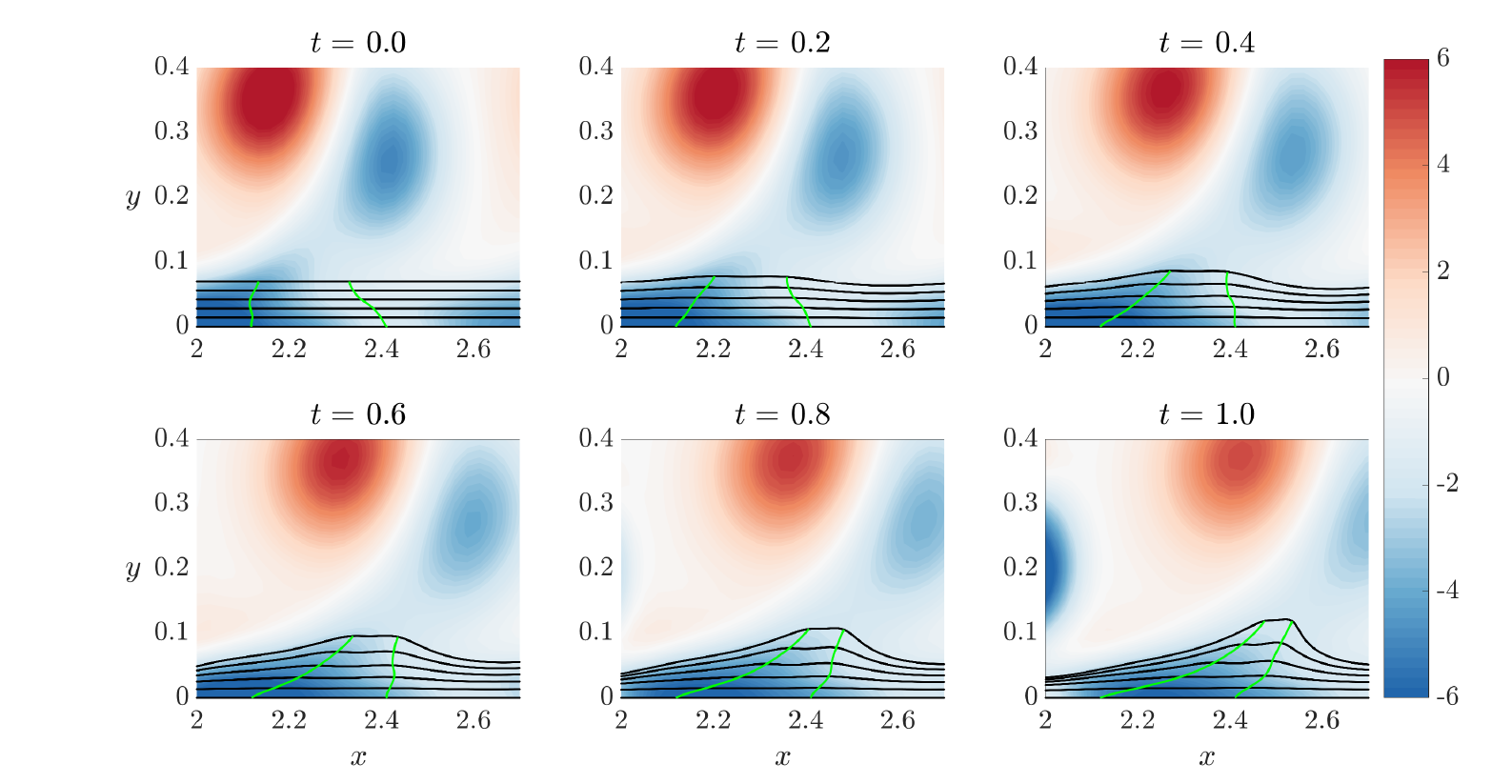}}
    \caption{Time evolution of the Lagrangian backbones $\B(t)$ of separation (green) computed from $\overline{\kappa}_0^1$, along with the vorticity field and material lines initially parallel to the wall (black) for the second vortex-induced separation in the impinging jet (see \fig{10}).} 
    \label{fig14}
\end{figure}

Similarly to \figs{11} and \ref{fig12}, \figs{13} and \ref{fig14} show the Lagrangian backbones of separation and their time evolution in correspondence of the second, vortex-induced separation indicated in \fig{10}. As in the previous case, there are two backbones, both connected to the wall, and contributing to the same long-term separation phenomenon. Levels of $\overline{\kappa}_0^1$, however, are two order of magnitude lower compared to \figs{11}, explaining the less sharp spike geometry, and a milder ejection of particles from the wall. The presence of less sharp spikes can be explained by the fact that as vortices travel radially, they are gradually ejected from the wall. This is better illustrated in \fig{14} where the centers of concentrated vorticity patches are located at a higher $y$-position compared with \fig{12}.

\Fig{15} shows the initial position of the Lagrangian backbone of separation extracted from a longer integration time $T=2$. The magnitude of $\overline{\kappa}$ shows that the downstream backbone is characterized by higher curvatures than the upstream backbone, in contrast to \fig{13} where the levels are comparable. More importantly, for $T=2$ the upstream backbone of separation in not connected to the wall. This suggests that for this longer integration time, the observed material spike is dominated by an off-wall separation process.

\begin{figure}[H]
    \centerline{\includegraphics[scale=.4]{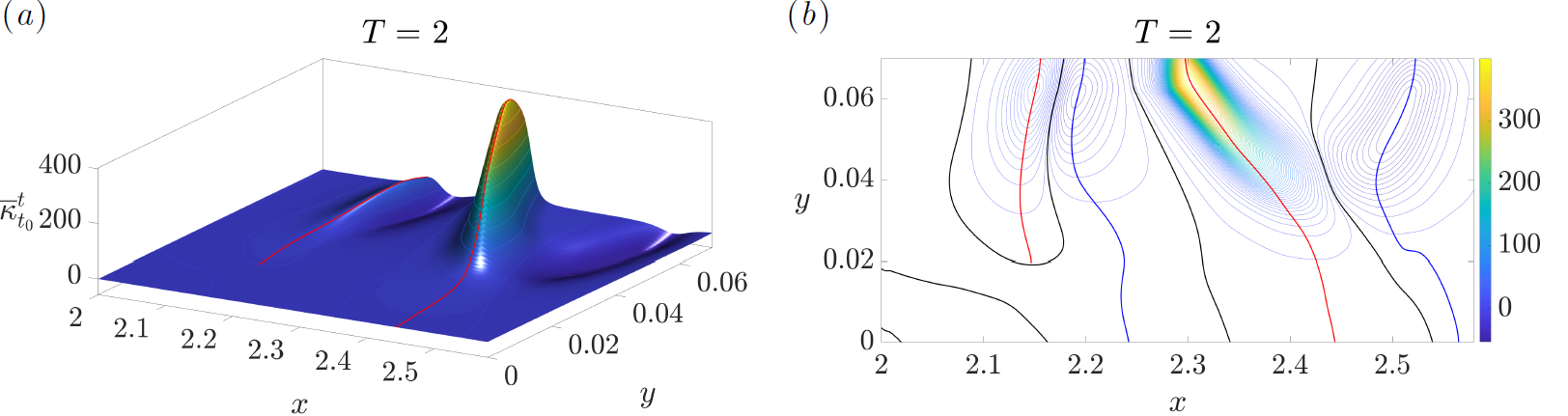}}
    \caption{Same as \fig{13} but for $T=2$.}
    \label{fig15}
\end{figure}

\begin{figure}[H]
    \centerline{\includegraphics[scale=1]{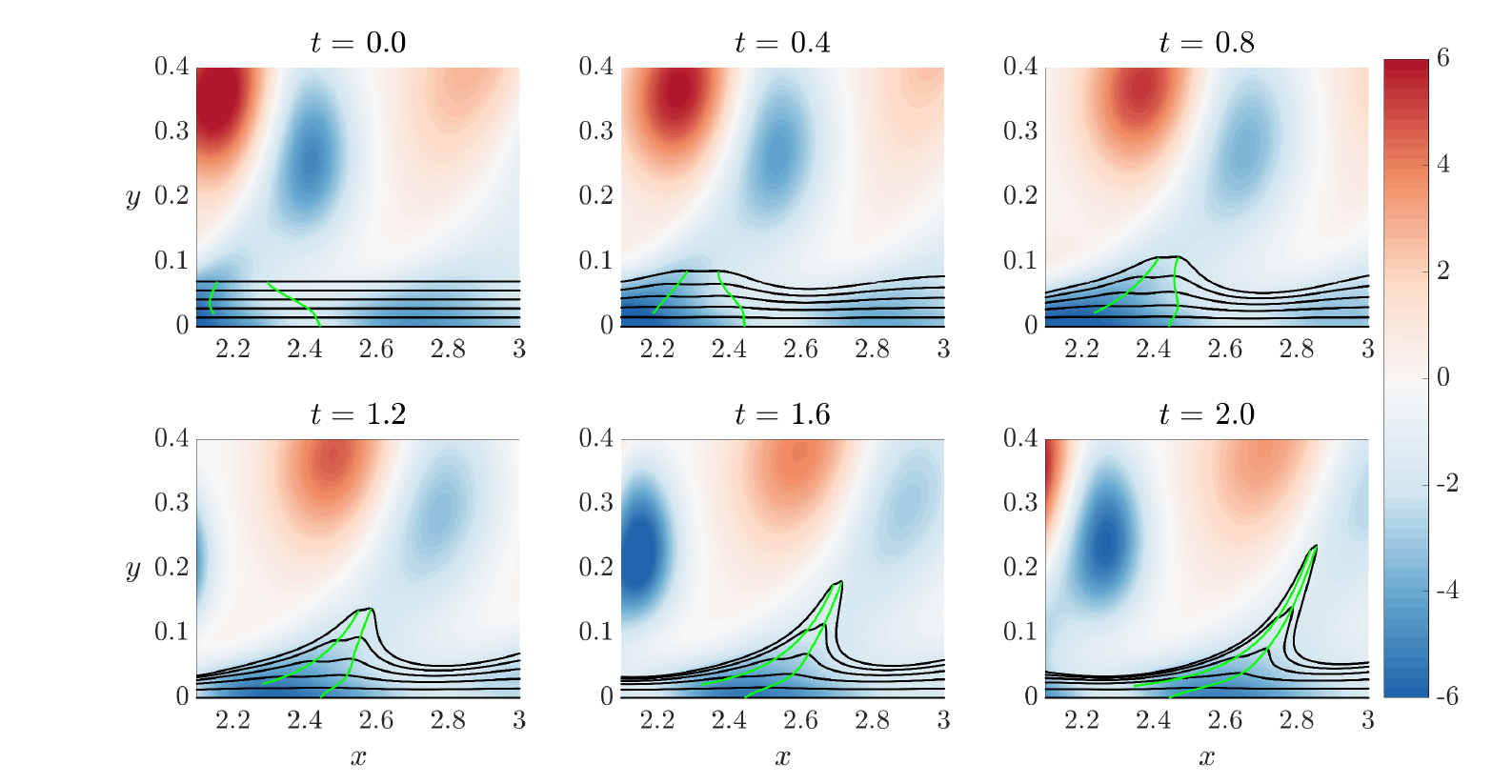}}
    \caption{Same as \fig{14} but for $T=2$.} 
    \label{fig16}
\end{figure}

This prediction is consistent with \fig{16}, which shows the evolving Lagrangian backbone $\B(t)$ at different times in green, along with evolving material lines initially parallel to the wall shown in black. For short time scales, the material spike has a footprint on the wall. As $T$ increases, vortices convect downstream decreasing their upwelling effect on particles close to the wall. This is clearly illustrated in \fig{16} as the $y$ location of the spike base point is constant in time. This result highlights that the theory of \cite{Serra2018} not only distinguishes on-wall and off-wall separation rigorously, but provides also the exact time scale at which a separation can switch from one type to the other. Such a transition is captured by a topological change in the $\kpmT$ field that remains hidden to previous theories.

Finally, we analyze the fixed separation illustrated in \fig{10}. \Fig{17} show the contour plots of the Lagrangian curvature change fields $\overline{\kappa}_0^6$ and $\overline{\kappa}_0^{12}$, along with the corresponding $\B(0)$. In contrast to \fig{13} and \ref{fig15}, here the backbones remain attached to the wall even for high integration times. Moreover, as the unsteady character of the flow in this region is weak, the spiking point is almost invariant with the integration time, consistent with the theoretical results in \cite{Serra2018}. \Fig{18} confirms that $\B(t)$ acts as the centerpiece of the forming spike, which remains completely hidden to the streamline geometry (blue) even in this quasi-steady flow region.
\begin{figure}[H]
    \centerline{\includegraphics[scale=1]{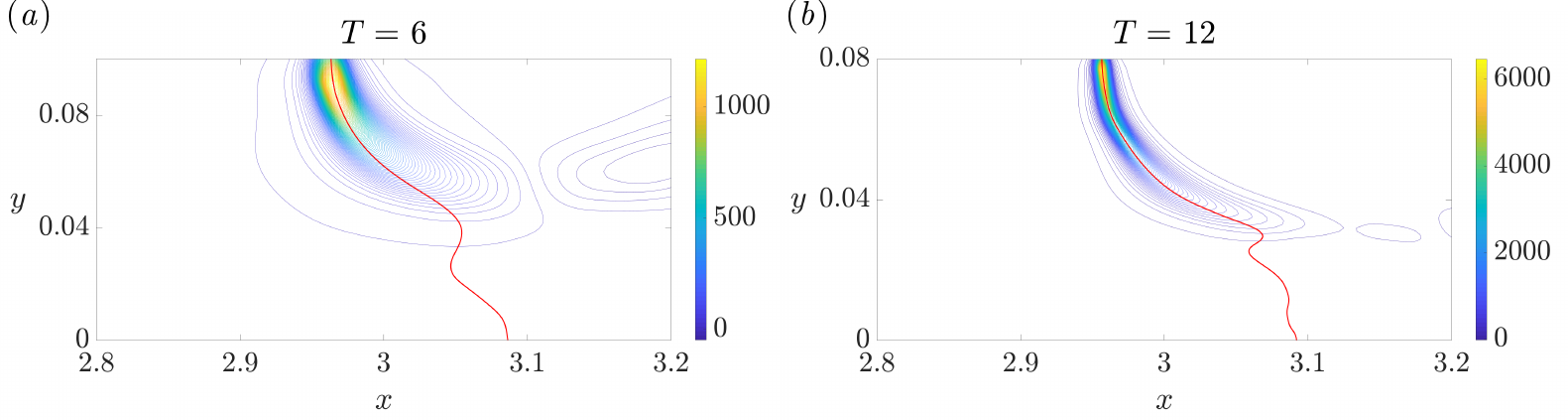}}
    \caption{Contour plot of the Lagrangian curvature change field $\kpmT$ for $t_0=0$, $T = 6$ \pa\ and $T = 12$ \pb\ for the fixed separation in the impinging jet (see \fig{10}).}
    \label{fig17}
\end{figure}

\begin{figure}[H]
    \centerline{\includegraphics[scale=1]{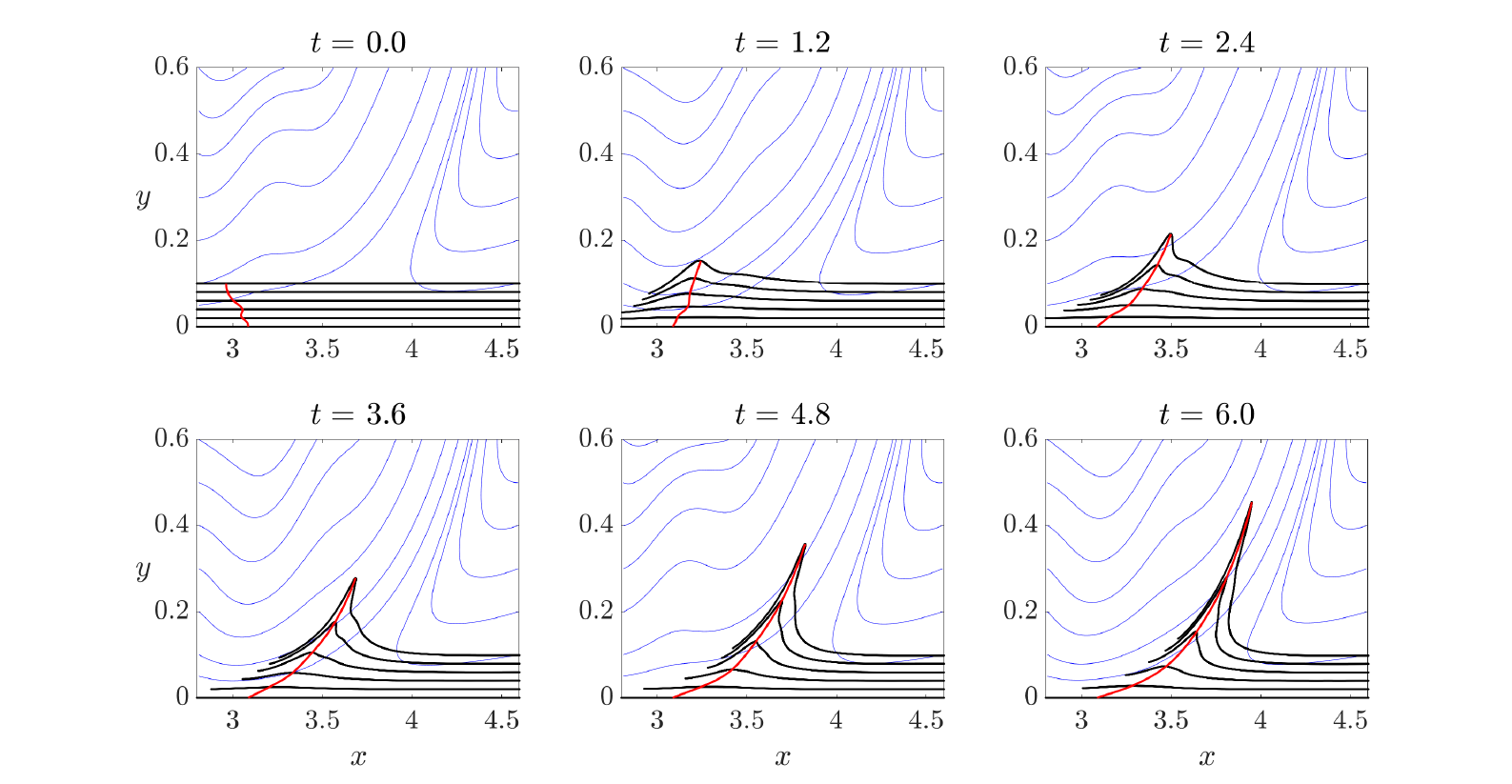}}
    \caption{Time evolution of the Lagrangian backbone of separation $\B(t)$ (red), extracted from $\overline{\kappa}_0^6$, along with streamlines (blue) and material lines initially parallel to the wall (black) for the fixed separation in the impinging jet (see \fig{10}).} 
    \label{fig18}
\end{figure}

\section{Flow around curved and moving boundaries}\label{fac}

In this section, we explore the phenomenon of spike formation in the vicinity of curved walls. We use a curvilinear coordinate system $(s,\eta)$ in a neighborhood of the boundary, as illustrated in Fig. \ref{fig:illustr}. 
In the case of moving walls, it is generally believed that separation takes place at a point off the boundary \citep[see, e.g.,][]{Sears1975}. For example, according to the Moore--Rott--Sears (MRS) principle \citep[after][]{Moore1958, Rott1956, Sears1956}, this point is located where the velocity and the wall-component of the shear vanishes. However, in \cite{Miron2015b}, it is shown that the MRS condition is not met in the flow generated by a translating cylinder rotating close to a fixed wall. The MRS principle can indeed be investigated only in the context of boundary layer theory, i.e. in the limit when $\Rey \rightarrow \infty$, again an asymptotic theory difficult to apply in practice. As an alternative, \cite{Miron2015b} proposed to detect separation as a Lagrangian saddle point identified by a distinguished location along an attracting LCS. In practice, this requires the extraction of attracting LCS, and then the identification of the point on this material line that maximizes the tangential rate of strain. \cite{Miron2015a} applied this technique to uncover the separation topology in a flow around a rotating cylinder facing a prescribed upstream velocity.

\begin{figure}[H]
    \centerline{\includegraphics[scale=1]{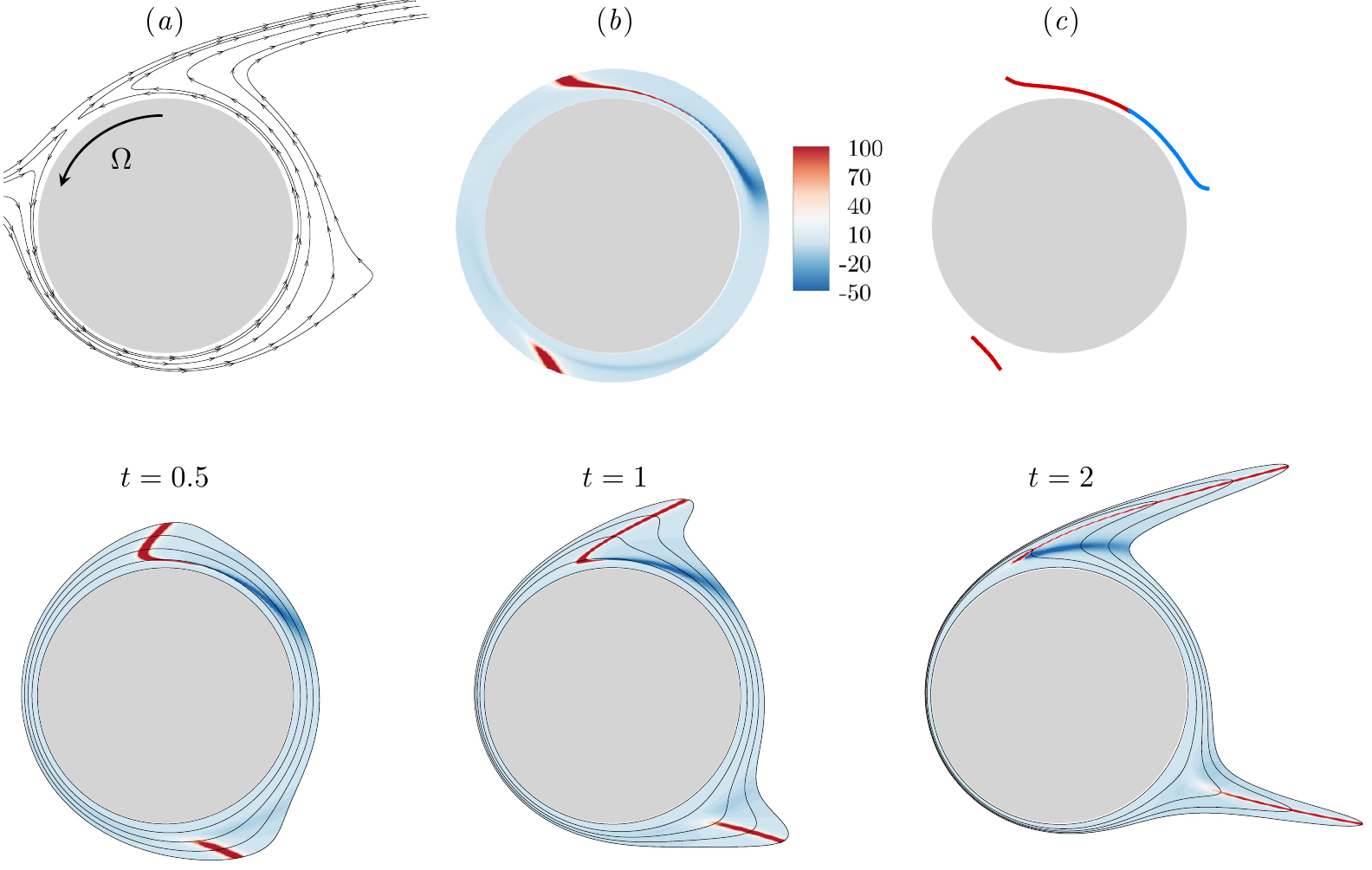}}
    \caption{Steady flow around a rotating cylinder at $\Rey = 50$ with $\Omega = 1$. \pa\ Streamlines, \pb\ contours of $\overline{\kappa}_0^2$ and \pc\ extraction of backbones of separation (red lines), together with the loci of minimal curvature in the upper region (blue line). The bottom figures show the materially advected curvature change field $\overline{\kappa}_0^2$ in \pb\ from $t=0$ to $t=2$ along with a set of material lines initially parallel to the cylinder.} 
    \label{fig19}
\end{figure}

In \Fig{19}, we show the spike formation in a steady flow arising around a circular cylinder rotating at a constant rotation rate $\Omega$, and immersed in a uniform flow characterized by $\Rey = 50$ (based on the inflow velocity and the cylinder diameter $D$). \Fig{19}\pa\ shows the streamlines around the cylinder obtained with $\Omega =1$. An Eulerian saddle point is present in the top left region of the cylinder. This point was detected by \cite{Miron2015a} as a distinguished point along an attracting LCS. Note that LCSs can be generally used in unsteady flows, as opposed to streamlines which are related to actual particle trajectories only in steady flows. \Fig{19}\pb\ shows the Lagrangian curvature change field $\overline{\kappa}_0^T$ for $T=2$. \Fig{19}\pc\ shows the initial position of the Lagrangian backbones of separation (red), along with the loci of minimal curvature (blue) for the top separation profile. Here separations are off-wall and backbones of separation end at about a distance of $0.03D$ from the wall.  The bottom plots of \fig{19} shows the materially advected curvature change field $\overline{\kappa}_0^2$ for three different times, along with the material lines initially parallel to the wall. The backbones of separation (red) act as centerpieces of the forming spikes, while the loci of minimal curvature (blue) precisely captures the sharp change in the spike geometry below the backbone.

While in \cite{Miron2015a} the Lagrangian saddle point was detected over a long time interval, \fig{19} illustrates that such a spike can be detected early on, and can form at a different location compared to the position of the long-term separated spike. As opposed to the upper separation, in the bottom region of the cylinder, no saddle point is present. In addition, neither the velocity nor the wall-component of the shear vanishes. Nevertheless, the $\overline{\kappa}_0^2$ field shows a strong ridge, i.e., a Lagrangian backbone of separation (\fig{19}\pc). As in the upper flow, the backbone is not connected to the wall, and captures the core of a separation spike as important as the upper one (\fig{19} bottom). It initiates upstream the cylinder, then moves with the cylinder by simultaneously ejecting fluid particles downstream. This separation, despite its severe intensity, was not detected by using LCSs techniques because of the absence of a Lagrangian saddle point. The present theory, instead, free from any a priori assumptions, promptly captures both the separation spikes. The case of the rotating cylinder is interesting as it includes simultaneously an upstream-moving (upper region) and a downstream-moving wall (lower region). In this last case in particular, the absence of reverse flow or recirculation region makes it difficult to detect separation with traditional approaches, explaining why the two separation phenomena are treated separately in the literature \citep[as for example by][]{Elliott1983}, contrary to the present study.

\subsection{Unsteady flow around a freely-moving cylinder}\label{cyl}

In this section, we consider the 2D flow around a circular cylinder of diameter $D$ that is fixed in rotation but free to translate in the axial $x$ and transverse $y$ directions. The Reynolds number based on the steady free stream velocity $U$ is set to $\Rey = UD/\nu = 100$. This dynamic fluid-structure simulation was carried out with an in-house finite element code. We solve the full Navier--Stokes equation simultaneously with an equation of motion governing the translation of the cylinder, which includes a structural stiffness and damping in both directions, and the contact force on the solid-fluid interface. This leads to a displacement of the cylinder with non-constant velocity and acceleration, thus providing a complex test case for our separation criterion. The reader is referred to {\cite{Gsell2016} for a detailed description of the simulation.

\begin{figure}[h!]
    \centerline{\includegraphics[scale=1]{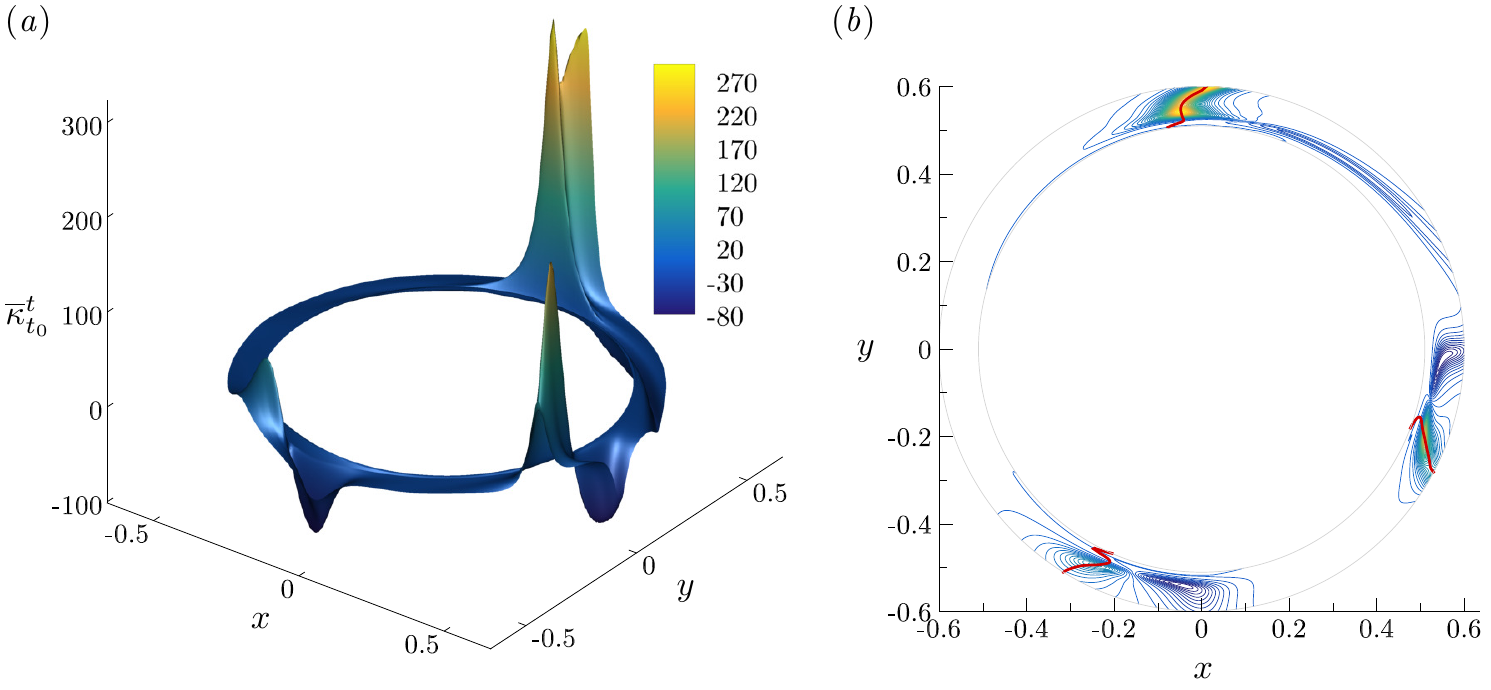}}
    \caption{\pa\ Lagrangian curvature change field $\overline{\kappa}_{0}^1$ around the cylinder free to move. \pb\ Contour plot of the scalar field in \pa, with Lagrangian backbones of separation (red lines).}
    \label{fig20}
\end{figure}

\Fig{20}\pa\ shows the surface plot of $\overline{\kappa}_{0}^1$ in the reference frame moving with the cylinder. The same $\kpmT$ would have been obtained, however, in any other reference frame, rotating and translating with respect to the cylinder, as $\kpmT$ is objective. Values of the Lagrangian curvature change field are particularly important on three distinct regions around the cylinder, the strongest separation being located at the top of the cylinder. By examining the backbones of the material spikes (red lines) in \fig{20}\pb\ superimposed on the contour plot of the scalar field in \fig{20}\pa, we find the three separation phenomena all originate from the wall as the three red lines are connected to the cylinder.

To better illustrate how fluid particles are ejected from the cylinder, \fig{21} shows the advection in time of $\overline{\kappa}_{0}^1$, detailed in \fig{20}, in an absolute frame of reference, thus allowing to visualize how the cylinder moves in time. Contours of the scalar field are plotted on a reduced scale to emphasize regions of positive (red) and negative (blue) curvature changes.
We note that while these contours do not apparently show the connection of backbones to the wall, in reality they are connected. This feature, for example, is illustrated in the multiple separation processes shown in \fig{5}. 
	
Starting from $x=0$, the second backbone of separation shows high curvature values from the interior of the flow down to the wall. In contrast, the three other separation processes show a significant drop of $\kpm$ from the flow interior to the wall, but backbones are nevertheless connected the no-slip boundary. This is also the case, for example, in \fig{1}\pc. A similar argument holds for the following figures.

\begin{figure}[H]
    \centerline{\includegraphics[scale=1]{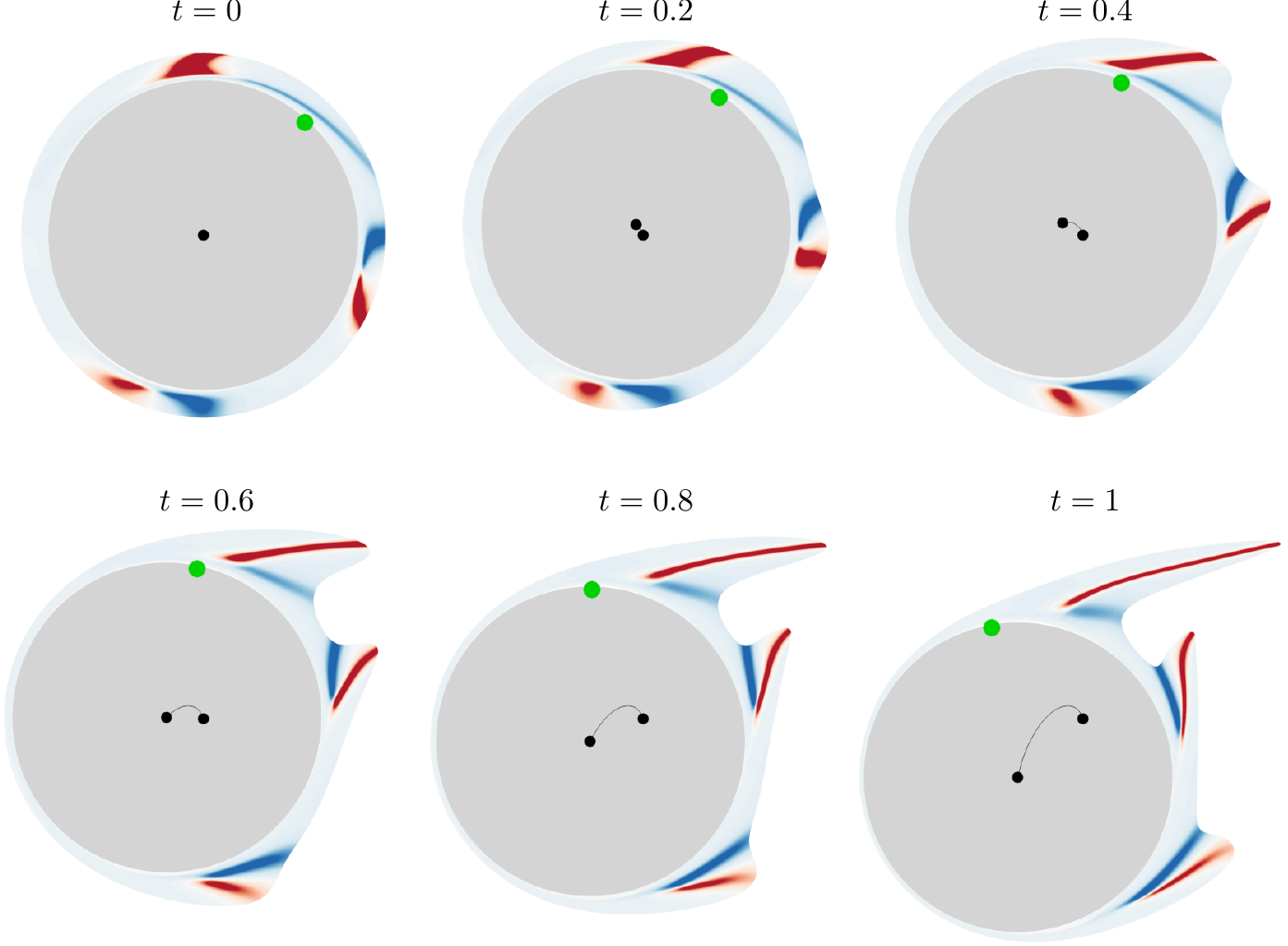}}
    \caption{Advection in time of the curvature change field shown in \fig{20}. Blue to red contours indicate negative to positive values of $\overline{\kappa}_{0}^1$. The figures are shown in a fixed frame of reference to visualize the cylinder's motion (black line), while the green dot marks the zero skin-friction point on the wall.} 
    \label{fig21}
\end{figure}
The cylinder goes up for $0<t<0.4$, and then down for $0.6<t<1$. Independently of this motion, the material advection of $\overline{\kappa}_{0}^1$ in time confirms the presence of three distinct Lagrangian separations, as the red contours coincide perfectly with the different material spikes. Additionally, the strong negative curvatures again captures sharp changes in the spike shape closed to the wall. Because of the curved wall geometry, a minimal curvature line appears only on one side of each Lagrangian backbone, as opposed to the flat wall cases discussed above. Finally, the green dots in \fig{21} indicate the Prandtl points. We observe that there is only one point where the wall shear-stress vanishes, clearly showing that the streamline pattern is totally uncorrelated to the separation processes in unsteady flows.

\Fig{22} is analogous to \fig{21} but corresponds to a different time interval $t\in [1,2]$, during which the cylinder accelerates towards the bottom direction. At the initial time $t_0=1$, $\overline{\kappa}_1^2$ is plotted over the undeformed set of material lines forming an annulus parallel to the wall, while at later times, the same scalar field is materially advected with the flow. Compared to the previous case (\fig{21}), two Lagrangian separation phenomena are detected instead of three, whereas we observe the presence of two Prandtl instead of one. As in \fig{21}, the Lagrangian curvature change field successfully capture two strong spikes ejecting fluid particles from the wall. One could wonder, however, how three separation phenomena turned into two. The reason is that around $t=1$, the two lowest separations in \fig{21} merge to form one single separation in \fig{22}. This will not be further detailed here, but will be more deeply investigated in the last example described in the next section.
\begin{figure}[H]
    \centerline{\includegraphics[scale=1]{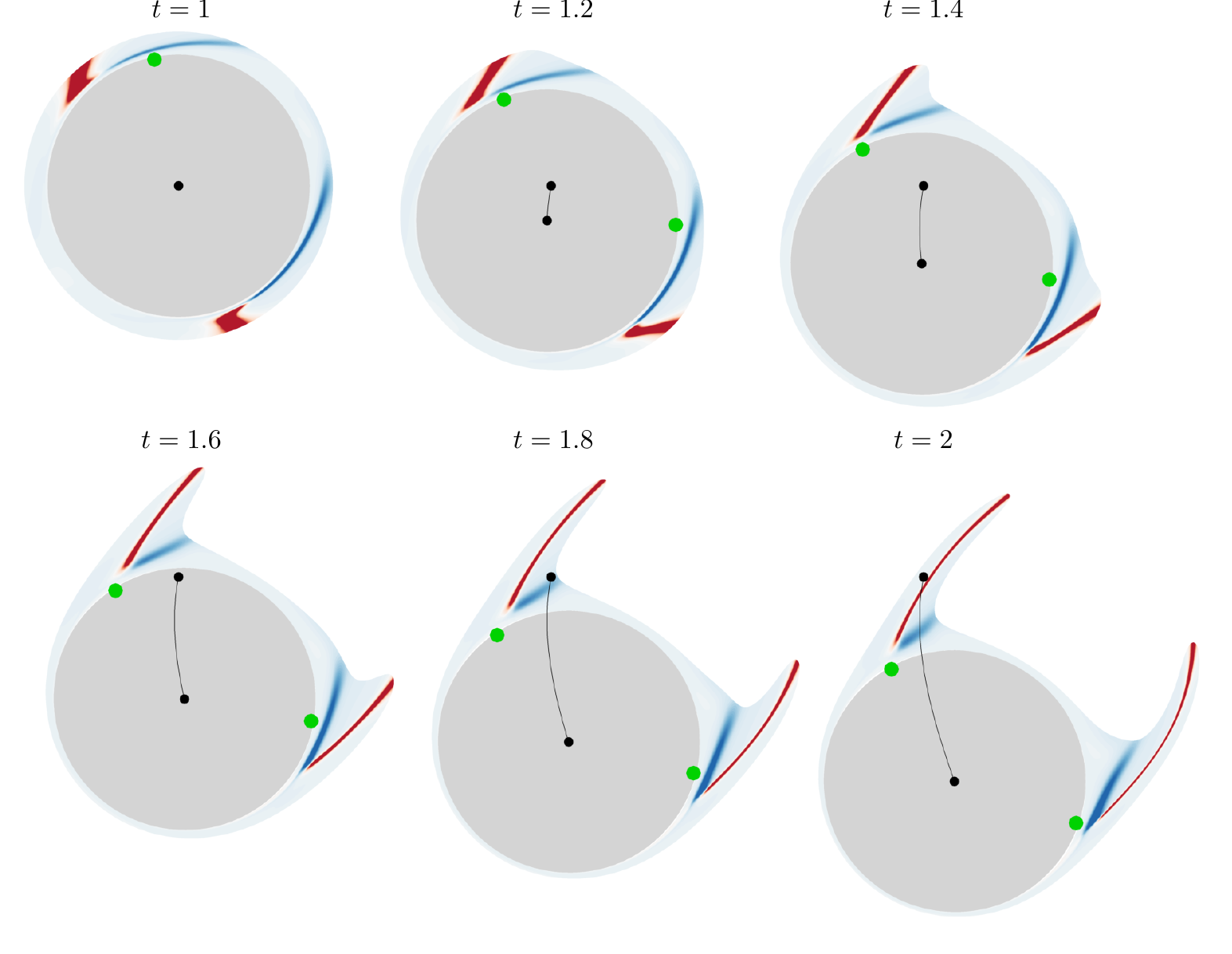}}
    \caption{Same as in \fig{21} but using $\overline{\kappa}_{1}^2$, and advecting it for $1\leq t\leq2$.} 
    \label{fig22}
\end{figure}

\subsection{Unsteady flow around a freely rotating ellipse}\label{ell}

Here we consider a two-dimensional flow around an elliptical cylinder that is free to rotate around a pivot point located upstream of the centroid of the ellipse at a distance of $D/4$ along the ellipse major axis. The major and minor axis lengths are $D=1$ and $d=0.5$. Denoting by $U$ the constant and uniform velocity of the free stream flow, the Reynolds number is $\Rey=UD/\nu =1000$. The fully coupled fluid-structure interaction simulation is similar to the case of the moving cylinder, except that the one degree of freedom angular equation of motion is solved together with the Navier--Stokes equation instead of the two degree of freedom translational motion. This relation includes also torsional structural stiffness and damping, and the torque relative to the pivot point arising from the contact forces on the solid-fluid interface. The numerical scheme is not the focus of this paper, but for further details on the simulations, performed with an in-house finite element code, the reader is referred to the similar work done by \cite{Weymouth2014}. Among most pertinent flow characteristics, the self-excited ellipse motion can be periodic, bistable, intermittently chaotic or fully chaotic, thus providing very rich databases for complex separation phenomena.

\Fig{30} shows the time evolution of the angle of rotation $\theta$ relative to the horizontal axis, in radians, and the angular velocity $\dot{\theta}$ of the ellipse, from the start of the simulation when the fluid is at rest, and the major axis of the ellipse is aligned with the horizontal axis. From $t=20$, the ellipse begins to rotate intermittently without apparent regularity, and accelerate or decelerate in the same manner. This represents a complex test case where separation phenomena are difficult to predict. We analyze two test cases corresponding to the time sequences indicated with dashed lines in \fig{30}. 
\begin{figure}[H]
    \centerline{\includegraphics[scale=1]{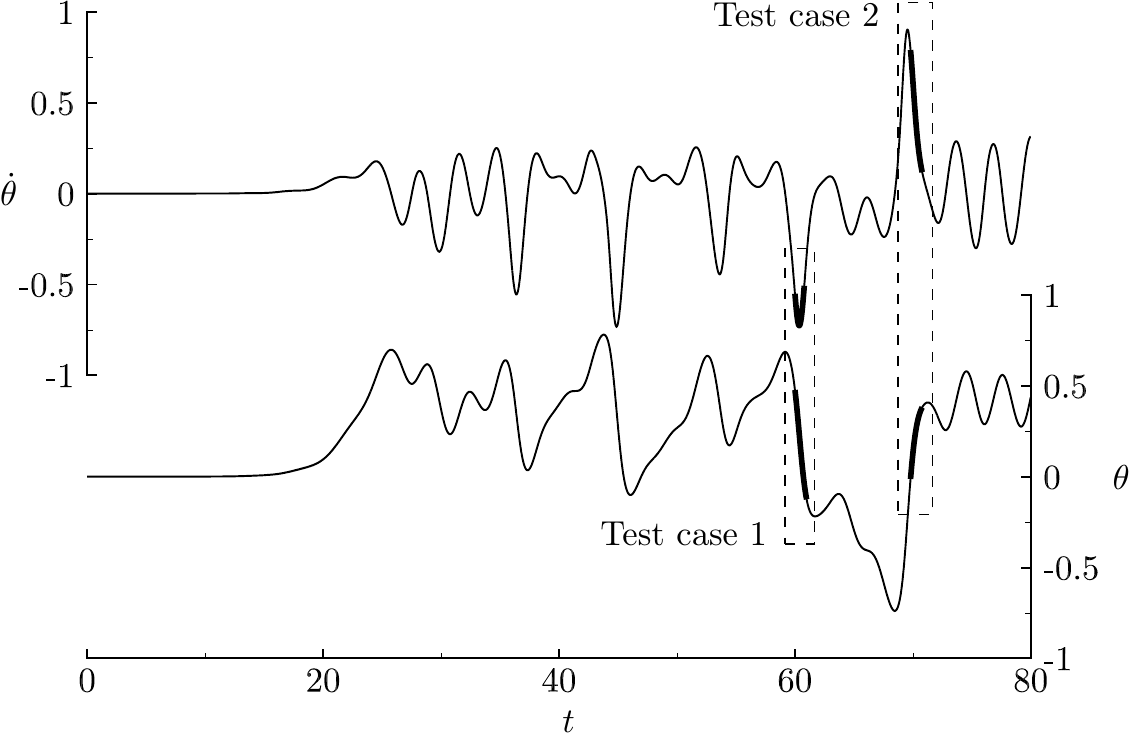}}
    \caption{Time history of the angle of rotation $\theta$ (bottom) and of the angular velocity $\dot{\theta}$ (top) defining the ellipse rotational motion about the pivot point.}
    \label{fig30}
\end{figure}

\begin{figure}[H]
    \centerline{\includegraphics[scale=1]{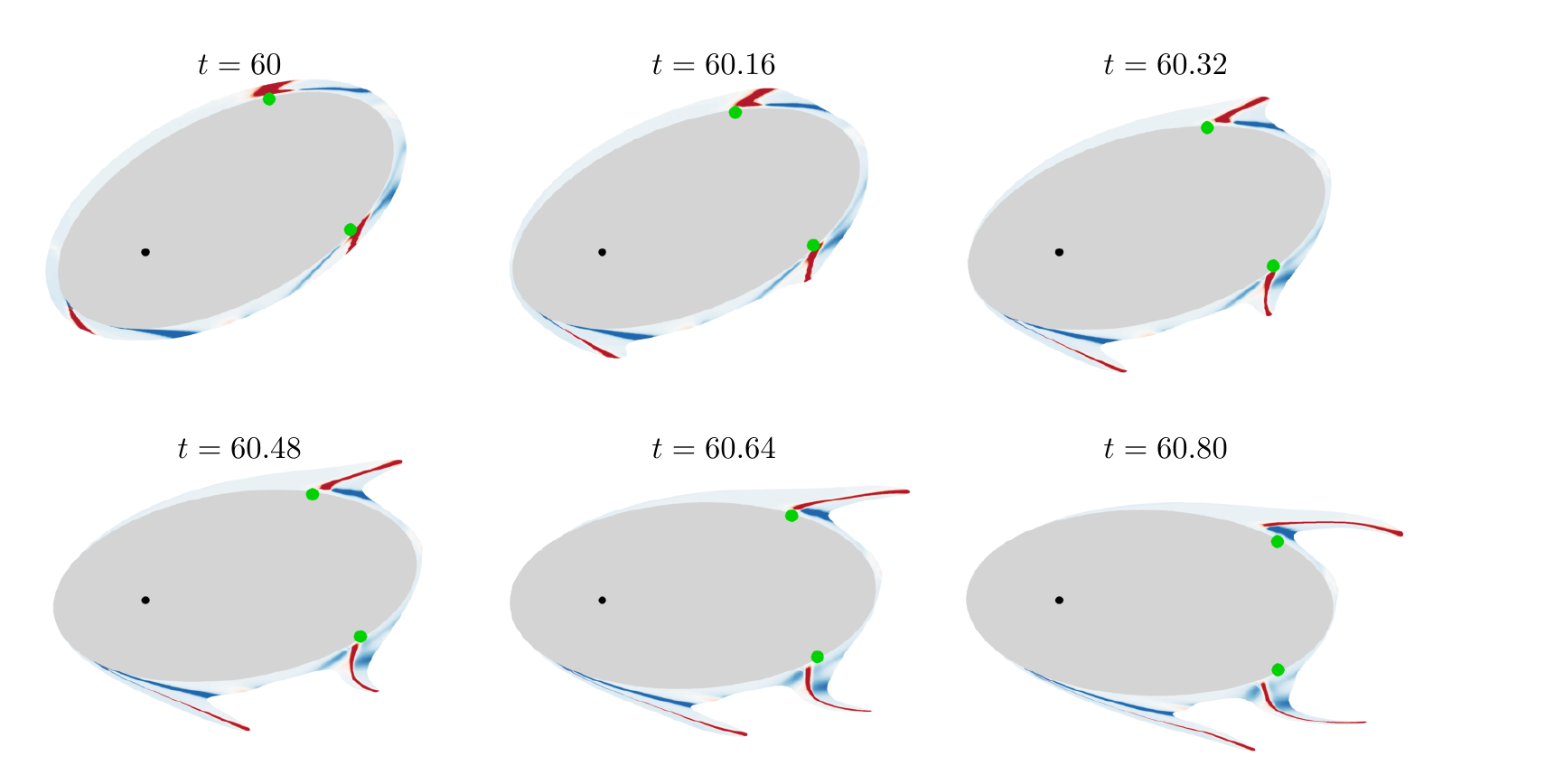}}
    \caption{ Test case 1 in \fig{30}. Lagrangian curvature change field $\overline{\kappa}_{60}^{60.08}$ plotted over the undeformed set of material lines initially parallel to the wall (top left), as well as on their advected positions at later times. Blue to red contours indicate negative to positive values of $\overline{\kappa}_{60}^{60.08}$. Green dots indicate the zero skin friction points.} 
    \label{fig31}
\end{figure}
\Fig{31} shows the Lagrangian curvature change field $\overline{\kappa}_{60}^{60.08}$, corresponding to test case 1 in \fig{30}. The top left panel  shows $\overline{\kappa}_{60}^{60.08}$ over the undeformed set of material lines initially parallel to the wall, while other panels show the advected $\overline{\kappa}_{60}^{60.08}$ at later times. For this case, the ellipse rotates in the clockwise direction, first with a phase of acceleration, then with a phase of deceleration. We detect three Lagrangian backbone of separation marked by red contours. By advecting the $\overline{\kappa}_{60}^{60.08}$ field until $t=60.8$, we again show how the separation backbones act as the centerpieces of the forming spikes. More generally, the $\kpmT$ field presents very rich informations of how the fluid particles in the close vicinity of the ellipse will move, deform, and eventually leave the surface. The Prandt points, are again unable to correctly capture the actual material spikes' location in unsteady flow separation.

\begin{figure}[H]
    \centerline{\includegraphics[scale=1]{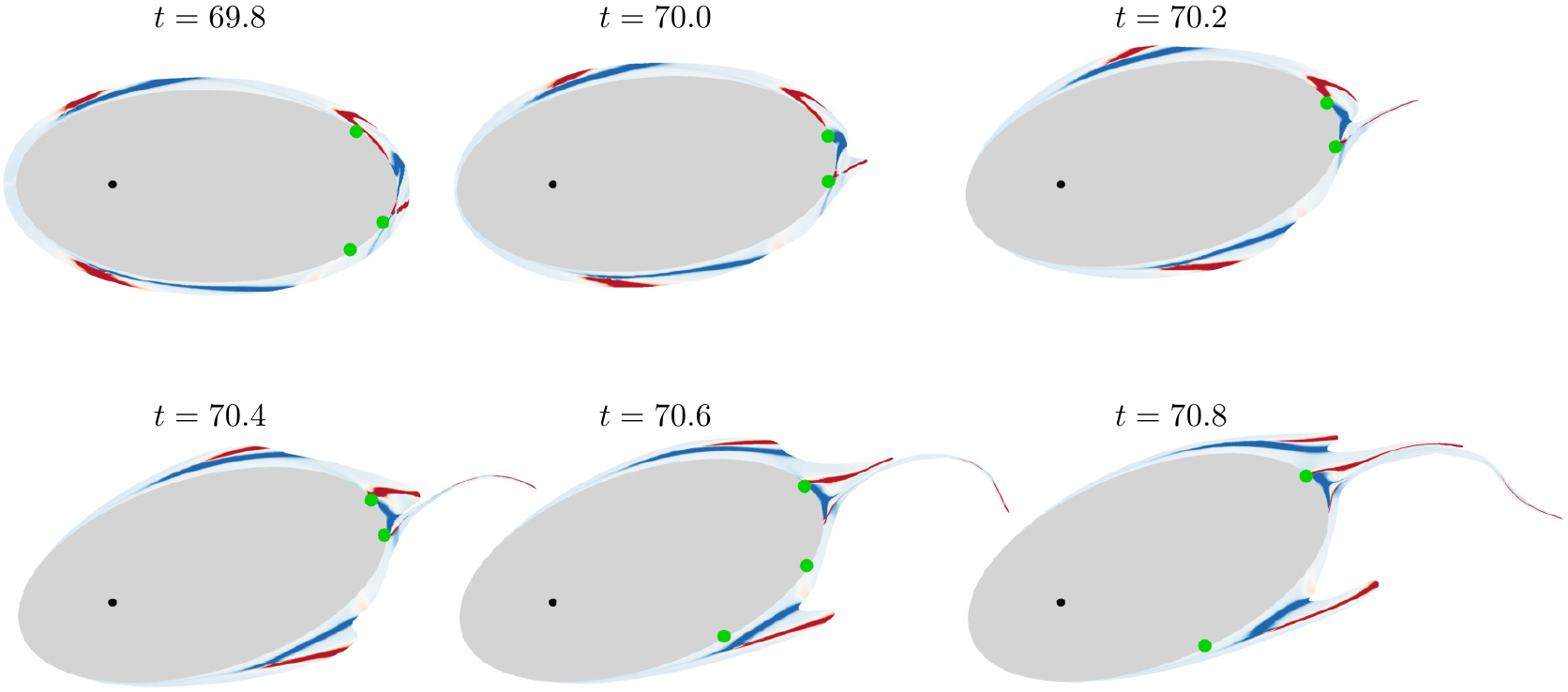}}
    \caption{Test case 2 in \fig{30}. Contours of the Lagrangian curvature change field $\overline{\kappa}_{69.8}^{70.8}$ at the initial time $t_0=69.8$, and its material advection at later times. Blue to red contours indicate negative to positive values of $\overline{\kappa}_{69.8}^{70.8}$. Green dots indicate the zero skin friction points.} 
    \label{fig32}
\end{figure}
\noindent \Fig{32} is analogous to \fig{31}, but corresponds to test case 2 in \fig{30}. In this scenario, the ellipse rotates counter-clockwise, strongly decelerating. One can observe that independently of the ellipse motion, separation locations are well-captured. In particular, we detect four initially distinct Lagrangian backbones of separation from $\overline{\kappa}_{69.8}^{70.8}$. At later times, however, the two rear-end backbone of separation gradually get closer to finally merge at the end of the time sequence. For this example, between one and three Prandtl points appear during the full evolution in time, showing again that the streamline pattern is not correlated to Lagrangian separation.

\Fig{33} shows other characteristic separation phenomena different from those presented previously in the flow around the ellipse. In \fig{33}\pab\, we show the Lagrangian curvature change fields $\kpmT$, over the corresponding final ($t=t_0+T$) fluid particle positions computed at two different initial times $t_0$. These two panels reveal the simultaneous development of four distinct and persistent separation spikes. In \fig{33}\pa\, the angle first increases from $t=t_0=63$ to $t=63.7$, then decreases until  $t=64.2$, thus experiencing a rotation inversion. \Fig{33}\pc\ shows yet another separation scenario illustrated through material advection of $\overline{\kappa}_{79}^{80.2}$ at three consecutive times. For $79<t<79.6$, we can observe the merge of the two top right backbones of separation into a unique separation spike, then for $79.6<t<80.2$, a new merge of this latter structure with the third separation spike developing in the rear-end of the ellipse. These test cases illustrate the great complexity of unsteady separation phenomena where several initially distinct material spikes can merge into a single common event, as precisely captured by the general theory of spike formation developed in \cite{Serra2018}.

\begin{figure}[H]
	\centerline{\includegraphics[scale=1]{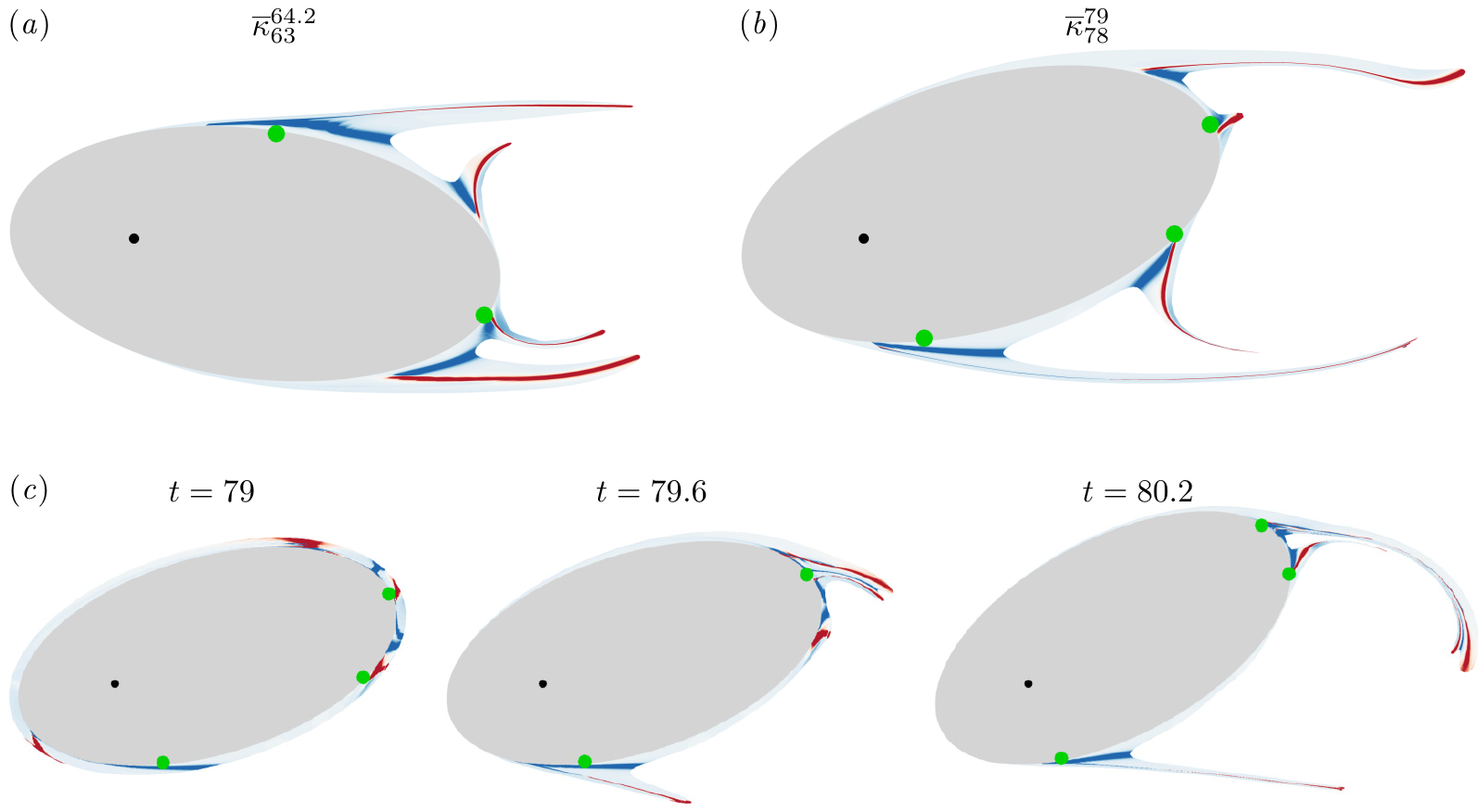}}
	\caption{Further examples of separation phenomena for the rotating ellipse. Panels \pa\ and \pb show $\kpmT$ over the final $t=t_0+T$ fluid particles position for two different $t_0$, where $t_0$ and $T$ are indicated in the legends. Panel \pc\ shows the time evolution of $\overline{\kappa}_{79}^{80.2}$. Green dots indicate the zero skin friction points.} 
	\label{fig33}
\end{figure}

\section{Conclusion}\label{con}

We have tested the theory of material spike formation during flow separation, as developed in \cite{Serra2018}, in several complex two-dimensional unsteady flows. These include the turbulent separation bubble, impinging jet, flow around a freely moving cylinder, and a freely rotating ellipse. In these flows, the theory has uncovered the Lagrangian and Eulerian backbones of separation. In contrast to prior asymptotic techniques, our approach has proven itself effective over short time scales, even instantaneously, capturing the separation spike from its birth to its fully developed Lagrangian shape. This is expected to provide a new tool for monitoring and controlling unsteady separation. 

The curvature-based theory of spike formation has also been able to detect the time scale over which on-wall separation transitions into off-wall separation (\S\,\ref{sec:IJRes}), as well as the merging processes of initially distinct separation spikes into a single feature (\S\S\,\ref{sec:SBres}, \ref{sec:IJRes} and \ref{ell}). Remarkably, we have found that even in simple flows, using existing techniques for identifying what was a priori believed to be off-wall separation, would miss well-pronounced separation spikes (Fig. \ref{fig19}) that we predict.
Based on our analysis here, the classic concept of `a' separation point appears too restrictive for highly unsteady flows, as separation can initiate on several locations, even locally. We have shown that our method is insensitive to this complexity, and applies to any surface shape, curved or not, moving or still, even when undergoes strongly accelerated motions. As future work, we plan to have a closer inspection of the topology of the Lagrangian curvature change field to quantify the stability of the different backbones of separation and predict which one is most likely to undergo an on/off wall transition or a merging process. It also appear worthwhile to explore the application of the same analysis in flow control strategies. In that context, one might consider a set of distributed actuators on the no-slip boundary, and design a control algorithm to optimally alter the flow to move the spiking points in a prescribed fashion. 
%

\section*{Acknowledgments}
M.S. would like to acknowledge the Schmidt Science Fellowship (\href{https://schmidtsciencefellows.org/}{https://schmidtsciencefellows.org/}). J.V. would like to gratefully acknowledge the financial support of the Natural Sciences and Engineering Research Council of Canada (NSERC), and also of the Simulation-based Engineering Science (Génie Par la Simulation) program funded through the CREATE program from the NSERC. This research was furthermore enabled by support provided by Calcul Qu\'ebec (\href{www.calculquebec.ca}{www.calculquebec.ca}) and Compute Canada (\href{www.computecanada.ca}{www.computecanada.ca}).

A special acknowledgment is dedicated to F\'elix, Danika and Pascal, students of Prof. A. Garon, who gratefully provided the databases on the moving cylinder and ellipse flows thanks to the in-house finite-element code they have developed.

\bibliographystyle{jfm}
\bibliography{jfm,temp,ReferenceList3_utf8,autres}

\end{document}